\documentclass[12pt]{iopart}

\usepackage{threeparttable}    
\usepackage{dcolumn}           
\newcolumntype{d}{D{.}{.}{-1}}
\usepackage{nomencl}           
\makeglossary
\usepackage{graphicx}          
\usepackage{caption}          
\usepackage{fancyvrb}          
\fvset{fontsize=\footnotesize,xleftmargin=2em}  
\usepackage{lettrine}          
\expandafter\let\csname equation*\endcsname\relax
\expandafter\let\csname endequation*\endcsname\relax
\usepackage{amsmath}           
\usepackage{hyperref}          
\graphicspath{{figs/}}         
\usepackage{float}             
\usepackage{upquote}
\usepackage{placeins}
\usepackage[retainorgcmds]{IEEEtrantools}
\usepackage{subfig}
\usepackage{siunitx}
\usepackage[export]{adjustbox}
\usepackage[title]{appendix}
\newcommand{\be}{\begin{equation}}
\newcommand{\ee}{\end{equation}} 

\usepackage{cite}
\usepackage[nameinlink,capitalise]{cleveref}

\begin{document}

\title[]{Multi-physics modeling of non-equilibrium phenomena in inductively coupled plasma discharges: Part II. Multi-temperature approach }

\author{Sanjeev Kumar, Alessandro Munaf\`{o}, Sung Min Jo, and Marco Panesi\footnote{\label{footnote_2}Corresponding author (mpanesi@illinois.edu).}  }

\address{Center for Hypersonics and Entry Systems Studies (CHESS), \\
Department of Aerospace Engineering, \\
University of Illinois at Urbana-Champaign, Urbana, IL 61801, USA}
\vspace{10pt}


\begin{abstract}
This paper provides a comparison between the vibrational-specific state-to-state (StS) model for nitrogen plasma elaborated in Part I of this work and conventional two-temperature (2-T) models for simulating inductively coupled plasma (ICP) discharges under non-Local Thermodynamic Equilibrium (NLTE) conditions. Simulations are performed within the multi-physics computational framework established for ICP in Part I. Based on the findings of Part I, the quasi-steady-state (QSS) assumption is validated in the plasma core, thereby enabling the calculation of global rate coefficients under this assumption. This facilitates the reduction of the StS model to a \textquotedblleft consistent\textquotedblright $\ $macroscopic 2-T model. Results from the StS model for nitrogen ICP torch exhibit considerable discrepancies when compared against predictions from the widely utilized Park 2-T model. On the contrary, the comparison between the newly proposed 2-T model, consistently derived from the original vibronic StS model, and the full StS results demonstrate excellent agreement in terms of plasma core location, morphology, and peak temperature distributions. This demonstrates the ability of the proposed 2-T model to capture the energy transfer and reactive processes predicted by the comprehensive StS model. Additionally, the study identifies the vibrational-translational (VT) energy transfer term in the 2-T model as the predominant factor in dictating plasma core morphology. This suggests a strong sensitivity of the ICP flow field to heavy-impact vibrational excitations and dissociative events.
\end{abstract}

\section{Introduction}\label{sec:intro}
Recent advances in both numerical and physical modeling have improved the accuracy of the predictions of inductively coupled plasmas (ICPs) \cite{Boulos_1976,mostaghimi1984parametric,mostaghimi1985analysis,mostaghimi1987two,proulx1987heating,mostaghimi1990effect,chen1991modeling}. These advancements have elucidated the intricate details of the thermochemical processes occurring under operating conditions pertinent to aerospace applications\cite{panesi2007analysis,abeele2000efficient,Kumar_RGD32,munafoRGD2022}. This manuscript represents Part II of a broader investigation: While Part I delves into the characterization of non-equilibrium effects in ICP plasma tunnels, Part II introduces a computationally efficient model designed to yield accurate results at a reduced computational cost. Traditionally, ICP simulations have relied on the assumption of Local Thermodynamic Equilibrium (LTE). Within this framework, the plasma state at any specific location is determined by its pressure and temperature, given a constant elemental fraction, which is achieved by maximizing the system's entropy\cite{reed1961induction,boulos1976flow,abeele2000efficient,utyuzhnikov2004simulation,colombo2010three,shigeta2012time,panesi2007analysis}. The LTE model's prevalent use stems from its computational efficiency and its applicability to thermal plasmas close to atmospheric-pressure conditions where LTE is predominantly observed. Yet, emerging research employing non-LTE simulations has demonstrated strong deviations from the equilibrium state for conditions of interest. Furthermore, these studies have also highlighted the profound influence of the chosen kinetic mechanism on the simulation outcomes\cite{mostaghimi1987two,Most_1989,panesi2007analysis,kumar2022self,kumar2022high,munafo2022multi,munafoRGD2022,el2007two}. 

In Part I of this work, using a coarse-grained vibrational StS model, we demonstrated severe deviations in the distribution from the Boltzmann equilibrium - a finding corroborated by other studies in the literature\cite{Kumar_RGD32,munafo2015tightly}. Similarly, discrepancies in the population distribution of vibrational levels from the Boltzmann distribution have been observed under comparable conditions in recombining nitrogen plasma experiments conducted at Stanford University's ICP facility \cite{gessman1997experimental,laux2012state}. Notably, this facility was moved to École Centrale Paris, where the scope of research expanded to encompass additional gas mixtures \cite{mcguire2022measurements,mcguire2020measurements,macdonald2015measurements}.

The most accurate modeling of NLTE plasma flows is achieved through the direct solution of the master equation, wherein each fundamental reactive process is distinctly modeled. This approach is known as the State-to-State (StS) method. StS models provide an accurate description of both collisional and radiative interactions across the internal energy levels of every species present in the flow\cite{laux2012state,bultel2013elaboration,munafo2015tightly,panesi2013rovibrational,colonna2015non,heritier2014energy,laporta2013electron,laporta2016electron,capitelli2013plasma,bultel2002influence, magin2006nonequilibrium, capitelli2007non,pietanza2010kinetic,munafo2012qct,munafo2013modeling,venturi2020bayesian,priyadarshini2022comprehensive,esposito1999quasiclassical,kustova2014chemical}.  By increasing the order of complexity and computational time, three primary variants of StS models can be defined: electronic \cite{jo2019electronic,panesi2009fire,panesi2011electronic}, vibrational \cite{elio_thesis,pereira2023vibronic}, and rovibrational state-to-state models \cite{panesi2013rovibrational,macdonald2018construction,macdonald2018construction2}. Electronic state-to-state models explicitly resolve the transitions between electronic states, with Boltzmann distributions for the remaining modes. On the other hand, vibrational state-to-state models account for the transitions between a molecule's vibrational states, defining only a rotational temperature. Both these models demand substantial computational resources, limiting their effective deployment in multidimensional codes. Consequently, most of the applications are constrained to 0-D and 1-D flow solvers. To address the inherent challenges of the StS model, recent innovations by the authors introduced a reduced-order technique. This approach, rooted in the coarse-graining method with the maximum entropy closure, offers a novel solution \cite{johnston2018impact,sahai2017adaptive,macdonald2018construction,macdonald2018construction2,munafo2014boltzmann,liu2015general,panesi2013collisional,venturi2020data,sahai2019flow,magin2012coarse,sharma2020coarse,kosareva2021four,zanardi2023adaptive,kuppa2023uncertainty}. Here, individual states of atoms and molecules are grouped into macro groups. The governing equations are then obtained by projecting the master equations onto this reduced subspace.

The analysis undertaken in Part I of this work has revealed that except for a narrow shell surrounding the plasma bubble, the overall discharge is in Quasi-Steady-State (QSS). In such conditions, the population of excited states can be deduced by solving a nonlinear algebraic system of equations. This method, which bypasses the need for time-dependent equations, offers a significant boost in computational efficiency. More importantly, the existence of a QSS distribution allows for the definition of macroscopic kinetic mechanisms that can be used within the framework of the multi-temperature models \cite{park1989nonequilibrium,park1993review,gnoffo1989conservation,yu2001effects,da2007two}. The 2-T model traditionally assumes equilibrium between the rotational and translational temperatures (T = Tr), and between free-electron, electronic and vibrational temperatures (Tv = Te). To calculate these temperatures and the energy exchanged between all the energy modes (i.e., translational, rotational, vibrational, and electronic), conservation equations for the internal energy modes in thermal nonequilibrium are added to the classical set of conservation equations for mass, momentum, and total energy. For the chemical kinetics model, macroscopic rate coefficients are assumed to depend on the different temperatures in the flow. It is important to mention that the values of these rate parameters suffer from large variability that can span orders of magnitude \cite{dunn1973theoretical,Gupta1990,park1993review,park2001chemical}. Also, the chemical-kinetic parameters (such as vibrational-translation (VT) coupling, chemistry-vibrational (CV) coupling, and other terms in the electro-vibrational energy equation) in the 2-T model have been found to influence the flow fields obtained from CFD calculations. It has been a common practice, at least in the hypersonics community, to use Park's chemical-kinetic parameters for the 2-T models in the CFD codes\cite{wright2009data,mazaheri2010laura,biedron2016fun3d,scalabrin2007numerical}. To overcome the inaccuracies of the chemical-kinetic parameters in the 2-T model, it is proposed to modify these chemical-kinetics parameters based on the state-to-state kinetic studies\cite{kim2018thermochemical, kim2020thermochemical,kim2021modification}, which gives more accurate results with 2-T model without resorting to computationally expensive StS models.

Hence, to overcome the limitation of the existing models, our objective is to devise a 2-T model that draws upon databases grounded in quantum mechanical analyses, a methodology previously utilized in Part I of this series. Furthermore, this paper presents a comparative analysis of the ICP results obtained from the nitrogen vibronic StS model (presented in Paper I) against the widely used 2-T models to assess the ability of the 2-T models to reproduce the StS results.

This paper is organized as follows: \cref{sec:physical_modeling} discusses the 2-T NLTE model for non-equilibrium plasma used in the present computational framework to describe the plasma inside the ICP facility. \cref{sec:reduction} discusses the strategy to reduce the vibronic StS model to a consistent 2-T model. \cref{sec:results} presents a comparative analysis of the ICP results obtained from the vibronic StS model against the one obtained from the widely used Park 2-T and the consistent 2-T (developed in this work) models. This section also presents a comparison of the ICP torch flow fields obtained from various physico-chemical models for a wide range of operating conditions to assess the applicability of various models given the ICP facility operating conditions. Finally, the conclusions are summarized in \cref{sec:conclusions}.

\section{Physical Modeling}\label{sec:physical_modeling}

The model for the electromagnetic field within the ICP torch is consistent with what was detailed in Paper I. However, in Paper I, the plasma description primarily revolved around the vibronic state-to-state assumption due to an emphasis on StS simulations. In this paper, we make use of the two-temperature (2-T) NLTE model, which will be utilized to compare with the vibronic StS findings.
\\

Under the assumptions listed in Part I, for a 2-T NLTE simulation, the plasma hydrodynamics are governed by the set of mass continuity, global momentum and energy, and vibronic energy equations \cite{Munafo_JCP_2020,Mitchner_book,gnoffo1989conservation}:
\begin{gather}
\frac{\partial \rho_{s}}{\partial t}+ {\nabla}_{\mathbf{r}}  \cdot\left[\rho_{s}\left(\mathbf{v}+ \mathbf{U}_s \right)\right] = \dot{\omega}_{s}, \quad s \in \mathcal{S}, \label{eq:cont} \\
\frac{\partial \rho \mathbf{v}}{\partial t}+ {\nabla}_{\mathbf{r}} \cdot(\rho \mathbf{v} \mathbf{v} + p \mathsf{I}) =  {\nabla}_{\mathbf{r}} \cdot \mathsf{\tau} + \mathbf{J} \times \mathbf{B}, \label{eq:momentum} \\
 \frac{\partial \rho E}{\partial t}+ {\nabla}_{\mathbf{r}} \cdot(\rho H \mathbf{v})= {\nabla}_{\mathbf{r}} \cdot \left( \mathsf{\tau} \mathbf{v} - \mathbf{q}\right) + \mathbf{J} \cdot \mathbf{E^{\prime}}, \label{eq:global_E} \\
\frac{\partial \rho e_{\mathrm{ve}}}{\partial t}+ {\nabla}_{\mathbf{r}} \cdot\left(\rho e_{\mathrm{ve}} \mathbf{v} \right)  =  - {\nabla}_{\mathbf{r}} \cdot \mathbf{q}_{\mathrm{ve}}  -p_{\mathrm{e}} {\nabla}_{\mathbf{r}} \cdot \mathbf{v} + \Omega_{\mathrm{ve}}^{\textsc{c}} + \mathbf{J} \cdot \mathbf{E^{\prime}},\label{eq:ve_eq}
\end{gather}

where $\mathcal{S}$ denotes the set of species, and the ve lower-script denotes the contributions for the sole \emph{vibronic} degrees of freedom. The various symbols in the governing equations \cref{eq:cont,eq:momentum,eq:global_E,eq:ve_eq} have their usual meaning: $t$ denotes time, $\mathbf{r}$ the position; $\rho$ and $\mathbf{v}$ the mass density and mass-averaged velocity, respectively; $\rho_s$ and $\mathbf{U}_s$ the partial density and diffusion velocity of species $s$; $p_{\mathrm{e}}$ the pressure of free-electrons; $e$ and $H$ the total energy and enthalpy per unit-mass, respectively; $\mathsf{\tau}$ the stress tensor; $\mathbf{q}$ the heat flux vector; $\dot{\omega}_s$ the mass production rates due to collisional processes; the $\Omega^{\textsc{c}}$ term the energy exchange terms due to collisional processes; $\mathbf{J}$ the conduction current density; $\mathbf{E}$ and $\mathbf{B}$ the electric field and the magnetic induction, respectively; $\mathbf{E^{\prime}} = \mathbf{E} + \mathbf{v} \times \mathbf{B}$ the electric field in the hydrodynamic frame (non-relativistic approximation). It is to be noted that the main difference in the plasma governing equations between Paper I and II lies in \cref{eq:ve_eq} which now represents the contribution from free-electron, electronic and vibrational modes instead of only free-electrons in Paper I. This leads to additional terms in the volumetric energy source term, $\Omega_{\mathrm{ve}}^{\textsc{c}}$. 
\\

The source term $\Omega_{\mathrm{ve}}^{\textsc{c}}$ in the electro-vibrational energy equation represents the energy exchange by elastic, inelastic and reactive collisional processes and is given by:
\begin{equation}
\Omega_{\mathrm{ve}}^{\textsc{c}}=\Omega^{\mathrm{VT}}+\Omega^{\mathrm{TE}}+\Omega^{\mathrm{DE}}+\Omega^{\mathrm{IE}}+\Omega^{\mathrm{CV}}+\Omega^{\mathrm{CEL}}
\end{equation}

$\Omega^{\mathrm{VT}}$ term denotes the inelastic vibrational energy exchange in atom-molecule and molecule-molecule collisions and is modeled based on the Landau-Teller model\cite{landau1936theorie}:
\begin{equation}
\Omega^{\mathrm{VT}}=\sum_{s \in \mathcal{S}_{\mathrm{m}}} \rho_s \frac{e_s^{\mathrm{V}}\left(T_{\mathrm{h}}\right)-e_s^{\mathrm{V}}\left(T_{\mathrm{e}}\right)}{\tau_s^{\mathrm{VT}}} \label{eq:omega_VT}
\end{equation}
The molecular VT relaxation times $\tau_S^{\mathrm{VT}}$ are computed as a frequency average of the atom-molecule and molecule-molecule VT relaxation times\cite{park1989nonequilibrium,gnoffo1989conservation} which are computed based on Millikan and White's formula\cite{millikan1963systematics} with Park's high-temperature correction\cite{park1993review} as:
\begin{equation}
\tau_{s r}^{\mathrm{VT}}=\frac{101325}{p} \exp \left[A_{s r}\left(T_{t r}^{-1 / 3}-B_{s r}\right)-18.42\right]
\end{equation}
where, $p$ is the pressure in Pascals and $T_{tr}$ is the trans-rotational temperature. The high-temperature correction is given as:
\begin{equation}
\tau_{c, s r}^{\mathrm{VT}}=\left(n_s \sigma_{v, s r} \sqrt{\frac{8 k_B T_{t r}}{\pi m_{r, s r}}}\right)^{-1}
\end{equation}
where $n_s$ is the number density of species $s$, and $\sigma_{v, s r}$ and $m_{r, s r}$ are the collision limiting cross-section and reduced mass between the species $s$ and $r$, respectively.

$\Omega^{\mathrm{CV}}$ denotes the volumetric rate of change of vibrational energy due to chemical reactions and is modeled by the non-preferential dissociation model\cite{candler1991computation}:
\begin{equation}
\Omega^{\mathrm{CV}}=\sum_{s \in \mathcal{S}_{\mathrm{m}}} \omega_s e_s^{\mathrm{v}}\left(T_{\mathrm{e}}\right)
\end{equation}
where, $e_s^{\mathrm{v}}\left(T_{\mathrm{e}}\right)$ denotes the average vibrational energy of species $s$ at temperature $T_{\mathrm{e}}$ computed using harmonic-oscillator assumption.

$\Omega^{\mathrm{CEL}}$ denotes the volumetric rate of change of electronic energy due to chemical reactions and is modeled by the non-preferential dissociation model:
\begin{equation}
\Omega^{\mathrm{CEL}}=\sum_{s \in \mathcal{S}_{\mathrm{m}}} \omega_s e_s^{\mathrm{el}}\left(T_{\mathrm{e}}\right)
\end{equation}
where, $e_s^{\mathrm{el}}\left(T_{\mathrm{e}}\right)$ denotes the average electronic energy of species $s$ at temperature $T_{\mathrm{e}}$.
\\

Definitions of $\Omega^{\mathrm{TE}}$, $\Omega^{\mathrm{DE}}$ and $\Omega^{\mathrm{IE}}$ have already been discussed in paper I. 
\\

The computation of the thermodynamic and transport properties remains the same as in Paper I.

\section{Reduction of the StS model to a consistent 2-T model}\label{sec:reduction}
    The flow field of the ICP is significantly influenced by the choice of reaction rates\cite{zhang2016analysis}. Furthermore, different kinetic mechanisms, such as Park and Dunn-Kang, may possess different reaction sets and even for the same reactions, the rates can depart significantly in magnitude. 
    This work aims to derive a reduced-order kinetic mechanism that is consistent with the StS model discussed in Part I. This allows us to assess the impact that the assumptions on the model form have on the results. Numerical integration of vibronic master equations is accomplished using the \textsc{plato}\cite{PLATO_AVIATION2023} library, applied across different bath temperatures. Global rate coefficients can subsequently be determined from the detailed state-specific rate coefficients combined with the vibronic state populations. The foundation for the global rates derives from earlier research \cite{panesi2013rovibrational,jo2022rovibrational,colonna2006reduction}, with the rates pertinent to the QSS region being selected for every reaction category. \cref{tab:macro_rates} presents the Arrhenius fit coefficients for the macroscopic rates, derived from the scaled-down vibronic StS kinetics rates. The formula for the Arrhenius form is as follows:

    \begin{equation}
    K = A T^B \exp \left(-\frac{C}{T}\right)
    \end{equation}
    where, $A$, $B$ and $C$ are the rate coefficients given in \cref{tab:macro_rates} and the units of $K$ are $m^3/mol.s$. 
    \\

    \begin{table}[hbt!]
    \caption{\label{tab:macro_rates} Arrhenius fit coefficients  $A$, $B$ and $C$ for the macroscopic rates obtained by reducing the vibronic StS kinetics rates}
    \small
    \centering
    \begin{tabular}{lcccc}
    \hline\hline
    Reaction & $A [m^3/mol.s]$ & $B$ & $C [K]$ \\
    \hline
    Heavy-impact dissociation\\
    $\mathrm{N}_2 + \mathrm{M} = \mathrm{N} + \mathrm{N} + \mathrm{M}$; $\mathrm{M} \in\left\{\mathrm{N},\mathrm{N}^{+}\right\}$ & $1.43\times10^{8}$ & 0.285 & $1.07\times10^{5}$ \\
    $\mathrm{N}_2 + \mathrm{M} = \mathrm{N} + \mathrm{N} + \mathrm{M}$; $\mathrm{M} \in\left\{\mathrm{N}_2,\mathrm{N}_2^{+}\right\}$ & $1.68\times10^{5}$ & 0.817 & $9.61\times10^{4}$ \\
    $\mathrm{N}_2^{+} + \mathrm{M} = \mathrm{N} + \mathrm{N}^{+} + \mathrm{M}$; $\mathrm{M} \in\left\{\mathrm{N},\mathrm{N}^{+}\right\}$ & $2.70\times10^{8}$ & 0.233 & $9.56\times10^{4}$ \\
    $\mathrm{N}_2^{+} + \mathrm{M} = \mathrm{N} + \mathrm{N}^{+} + \mathrm{M}$; $\mathrm{M} \in\left\{\mathrm{N}_2,\mathrm{N}_2^{+}\right\}$ & $1.55\times10^{10}$ & -0.246 & $9.85\times10^{4}$ \\
    \\
    \hline
    Electron-impact dissociation\\
    $\mathrm{N}_2 + \mathrm{e}^- = \mathrm{N} + \mathrm{N} + \mathrm{e}^-$ & $9.61\times10^{12}$ & -0.719 & $1.13\times10^{5}$ \\
    \\
    \hline
    Electron-impact ionization\\
    $\mathrm{N} + \mathrm{e}^- = \mathrm{N}^+ + \mathrm{e}^- + \mathrm{e}^-$  & $1.50\times10^{16}$ & -0.980 & $1.53\times10^{5}$ \\
    $\mathrm{N}_2 + \mathrm{e}^- = \mathrm{N}_2^+ + \mathrm{e}^- + \mathrm{e}^-$  & $1.05\times10^{3}$ & 1.556 & $1.75\times10^{5}$ \\
    \\
    \hline
    Dissociative recombination\\
    $\mathrm{N}_2^{+} + \mathrm{e}^- = \mathrm{N} + \mathrm{N}$ & $1.26\times10^{17}$ & -1.70 & $2.46\times10^{3}$\\
    \\
    \hline
    Charge exchange\\
    $\mathrm{N}_2 + \mathrm{N}^+ = \mathrm{N} + \mathrm{N}_2^{+}$ & $3.99\times10^{-3}$ & 2.190 & $2.10\times10^{3}$\\
    \\
    \hline
    Heavy-impact ionization\\
    $\mathrm{N} + \mathrm{N} = \mathrm{N}^+ + \mathrm{N} + \mathrm{e}^-$  & $1.28\times10^{21}$ & -2.593 & $1.96\times10^{5}$ \\
    $\mathrm{N} + \mathrm{N}_2 = \mathrm{N}^+ + \mathrm{N}_2 + \mathrm{e}^-$  & $4.48\times10^{20}$ & -2.516 & $1.93\times10^{5}$ \\
    \hline\hline
    
    \end{tabular}
    \end{table}

\begin{figure}[!htb]
        \centering
        \includegraphics[scale=0.2]{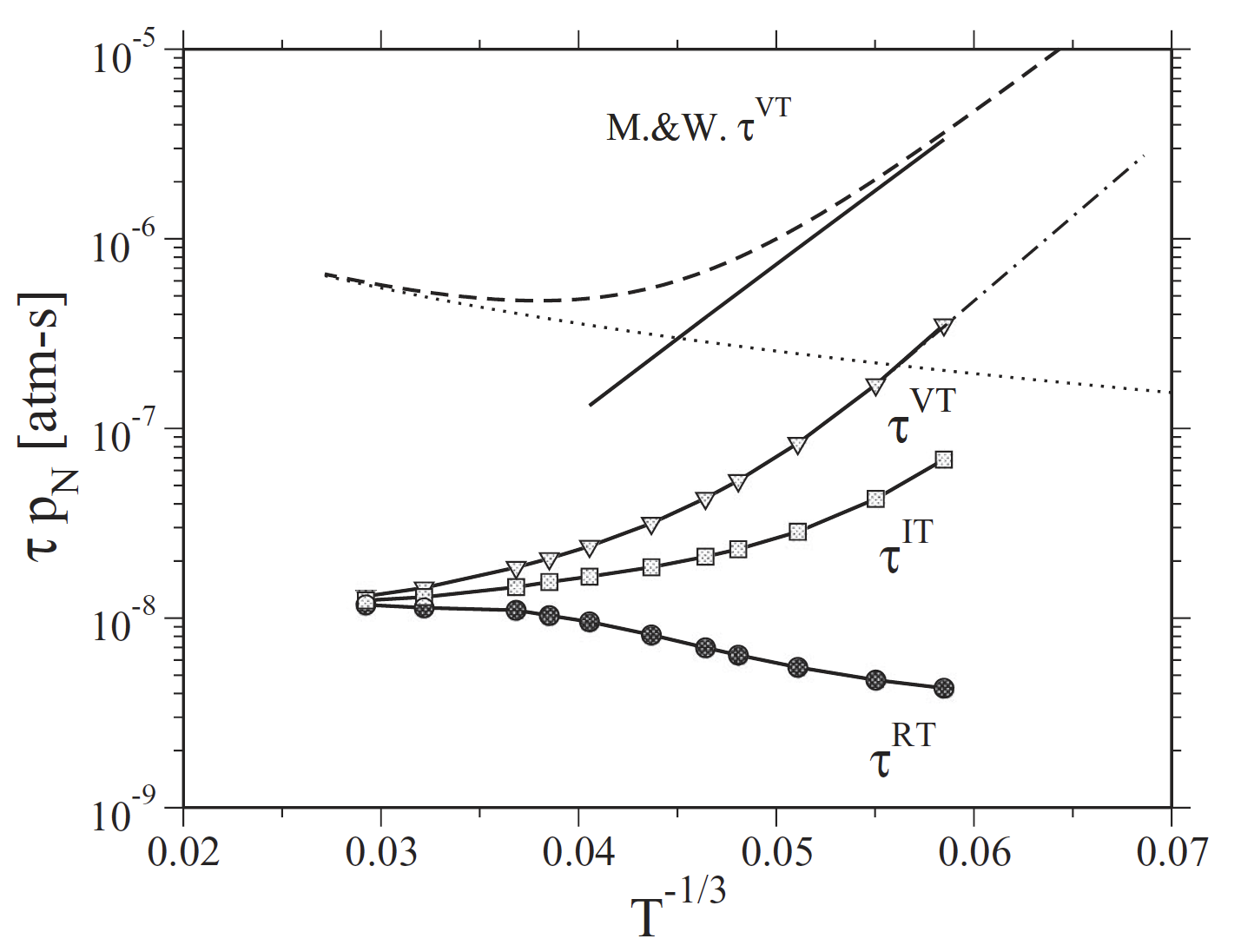}
        \caption{Comparison of relaxation times obtained from state-to-state kinetics against Millikan-White correlation formula for N\textsubscript{2}-N system (image taken from \cite{panesi2013rovibrational} with permission). }
        \label{fig:tauVT}
    \end{figure}

 To make the 2-T model fully consistent with the vibronic StS model, the vibrational relaxation time ${\tau_s}^{VT}$ in \cref{eq:omega_VT} should be computed based on the vibronic StS kinetics itself as recent state-to-state kinetic calculations of the atmospheric gas species found that Park's high-temperature corrected vibrational relaxation time does not agree with the one calculated using the state-to-state kinetics\cite{kim2013state,kim2021modification,andrienko2018vibrational,grover2019vibrational,macdonald2018construction,macdonald2018construction2,kim2018thermochemical,kim2020thermochemical}. \cref{fig:tauVT} shows a comparison of the vibrational relaxation times obtained from rovibrational state-to-state kinetics with the one obtained using MW + high-temperature correction formula for N\textsubscript{2}-N system\cite{panesi2013rovibrational}, depicting at least an order of magnitude difference. Hence, it is important to use vibrational relaxation times consistent with the state-to-state kinetics in order to get a good agreement with StS results. In the present work, the vibrational relaxation times for each electronic level of the molecules are evaluated based on the e-folding method\cite{park2004rotational}. The global relaxation time for a given molecule is then computed by taking a Boltzmann weighted sum of the relaxation times of all the electronic states as 
    \begin{equation}
    \frac{1}{\tau_{\text {global }}\left(T, p_{\mathrm{A}}\right)}=\frac{1}{\sum_l Q_l} \sum_n\left(\frac{Q_n}{\tau_n\left(T, p_{\mathrm{A}}\right)}\right), l, n \in I_{\mathrm{El}}
    \end{equation}
    where $I_{\mathrm{El}}=[0,1,2, \ldots]$ represents the set of electronic levels of the molecule and $Q_l$ denotes the electronic partition function. The global vibrational relaxation times are then fitted into the default MW + high-temperature correction expression as tabulated in \cref{tab:tauVT}. For the $\mathrm{N}_2^{+} - \mathrm{M}$ systems, same fits as $\mathrm{N}_2 - \mathrm{M}$ systems have been used in this work. 
    \\

    \begin{table}[hbt!]
    \caption{\label{tab:tauVT} Consistent parameters $A_{sr}$, $B_{sr}$ and $\sigma_{sr}$ for the vibrational relaxation times}
    \centering
    \begin{tabular}{lcccc}
    \hline\hline
    Type of collision & $A_{sr}$ & $B_{sr}$ & $\sigma_{sr}$ \\
    \hline
    N\textsubscript{2}+N & 184.44 & 0.0389 & $1.0 \times 10^{-22} T_{t r}^{0.52}$ \\
    N\textsubscript{2}+N\textsuperscript{+} & 184.44 & 0.0389 & $1.0 \times 10^{-22} T_{t r}^{0.52}$ \\
    N\textsubscript{2}+N\textsubscript{2} & 221.53 & 0.0290 & $1.7 \times 10^{-23} T_{t r}^{0.55}$ \\
    N\textsubscript{2}+$\mathrm{N}_2^+$ & 221.53 & 0.0290 &  $1.7 \times 10^{-23} T_{t r}^{0.55}$ \\
    $\mathrm{N}_2^+$+N & 184.44 & 0.0389 & $1.0 \times 10^{-22} T_{t r}^{0.52}$ \\
    $\mathrm{N}_2^+$+N\textsuperscript{+} & 184.44 & 0.0389 & $1.0 \times 10^{-22} T_{t r}^{0.52}$ \\
    $\mathrm{N}_2^+$+N\textsubscript{2} & 221.53 & 0.0290 & $1.7 \times 10^{-23} T_{t r}^{0.55}$ \\
    $\mathrm{N}_2^+$+$\mathrm{N}_2^+$ & 221.53 & 0.0290 &  $1.7 \times 10^{-23} 
 T_{t r}^{0.55}$ \\ 
    \hline\hline
    
    \end{tabular}
    \end{table}

\begin{figure}[!htb]
        \centering
        \includegraphics[scale=0.75]{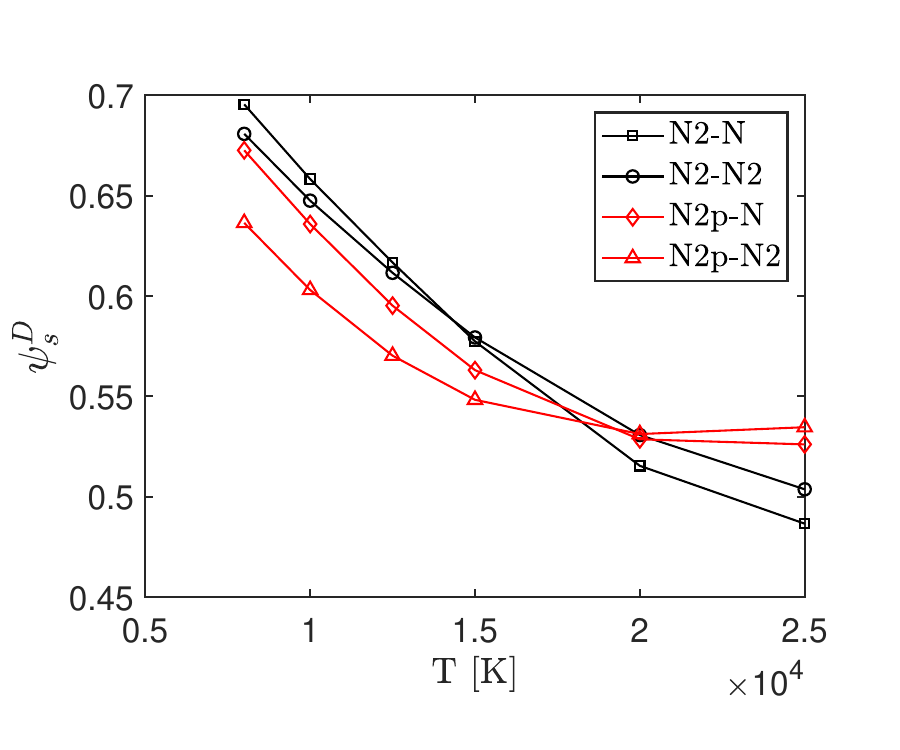}
        \caption{Vibrational energy removal ratio obtained from the vibrational StS kinetics}
        \label{fig:CV}
    \end{figure}

 Next, to make the 2-T model even more consistent with the vibronic StS model, the preferential dissociation model\cite{park1989assessment,park1993review,park2001chemical} is used to describe the volumetric change of vibrational energy due to chemical reactions instead of the non-preferential dissociation model used with the Park 2-T model in this work. With the preferential dissociation model, $\Omega^{CV}$ term in the electro-vibrational energy can be written as:
    \begin{equation}        
    \Omega^{CV}=\sum_{s \in \mathcal{S}_{\mathrm{m}}} \Dot{\omega}
     _s \psi_s^D E_s^D 
    \end{equation}
    where $\psi_s^D$ and $E_s^D$ are vibrational energy loss ratio and average dissociation energy of a given molecule. The definition of vibrational energy loss ratio is taken from Panesi et al\cite{panesi2013rovibrational}. To compute $\psi_s^D$, the vibronic master equations are again numerically integrated in time for a 0D isochoric reactor for various fixed bath temperatures and the values corresponding to the QSS region are selected as most of the energy transfer takes place in that region. \cref{fig:CV} shows the variation of $\psi_s^D$ with temperature for all the interactions, which is different from a constant value of 0.3 as proposed by Park \emph{et al}\cite{park1993review,park2001chemical}. The vibrational energy loss ratio is then fitted as a function\cite{kim2020thermochemical}:
    \begin{equation}
    \psi_{s}^D=\exp \left(K_1 / T+K_2+K_3 \ln \left(T\right)+K_4 T+K_5 T^2\right)
    \end{equation}
    \cref{tab:CV} tabulates the parameters $K_1$ to $K_5$ for the vibrational energy loss ratio obtained from state-to-state kinetics. 
    \\
    
    The vibronic StS reduced 2-T model with consistent $\tau_{VT}$ and $\psi_s^D$ will be referred to as \textquotedblleft consistent 2-T model\textquotedblright $\ $in this paper to differentiate from the Park 2-T model.

    \begin{table}[hbt!]
    \caption{\label{tab:CV} Parameters $K_1$ to $K_5$ for the vibrational energy loss ratio}
    \centering
    \begin{tabular}{lcccccccc}
    \hline\hline
    Reaction & $K_1$ & $K_2$ & $K_3$ & $K_4$ & $K_5$ & Min & Max \\
    \hline
    N\textsubscript{2}+N & 1.04E+04 & -2.39E+01 & 2.68E+00 & -2.54E-04 & 3.20E-09 & 0.486 & 0.813 \\
    N\textsubscript{2}+N\textsuperscript{+} & 1.04E+04 & -2.39E+01 & 2.68E+00 & -2.54E-04 & 3.20E-09 & 0.486 & 0.813 \\
    N\textsubscript{2}+N\textsubscript{2} & 3.29E+03 & -7.77E+00 & 8.59E-01 & -1.05E-04 & 1.39E-09 & 0.503 & 0.754  \\
    N\textsubscript{2}+$\mathrm{N}_2^+$ & 3.29E+03 & -7.77E+00 & 8.59E-01 & -1.05E-04 & 1.39E-09 & 0.503 & 0.754  \\
    $\mathrm{N}_2^+$+N & -4.62E+03 & 7.71E+00 & -8.39E-01 & -4.99E-06 & 7.17E-10 & 0.526 & 0.696 \\
    $\mathrm{N}_2^+$+N\textsuperscript{+} & -4.62E+03 & 7.71E+00 & -8.39E-01 & -4.99E-06 & 7.17E-10 & 0.526 & 0.696 \\
    $\mathrm{N}_2^+$+N\textsubscript{2} & -8.50E+03 & 1.83E+01 & -2.06E+00 & 1.10E-04 & -7.48E-10 & 0.534 & 0.654 \\
    $\mathrm{N}_2^+$+$\mathrm{N}_2^+$ & -8.50E+03 & 1.83E+01 & -2.06E+00 & 1.10E-04 & -7.48E-10 & 0.534 & 0.654 \\ 
    \hline\hline

    \end{tabular}
    \end{table}


\section{Results}\label{sec:results}

    \subsection{Problem description}\label{sec:problem_description}
\begin{figure}[!ht]
\centering
\subfloat[][]{\includegraphics[scale=0.25]{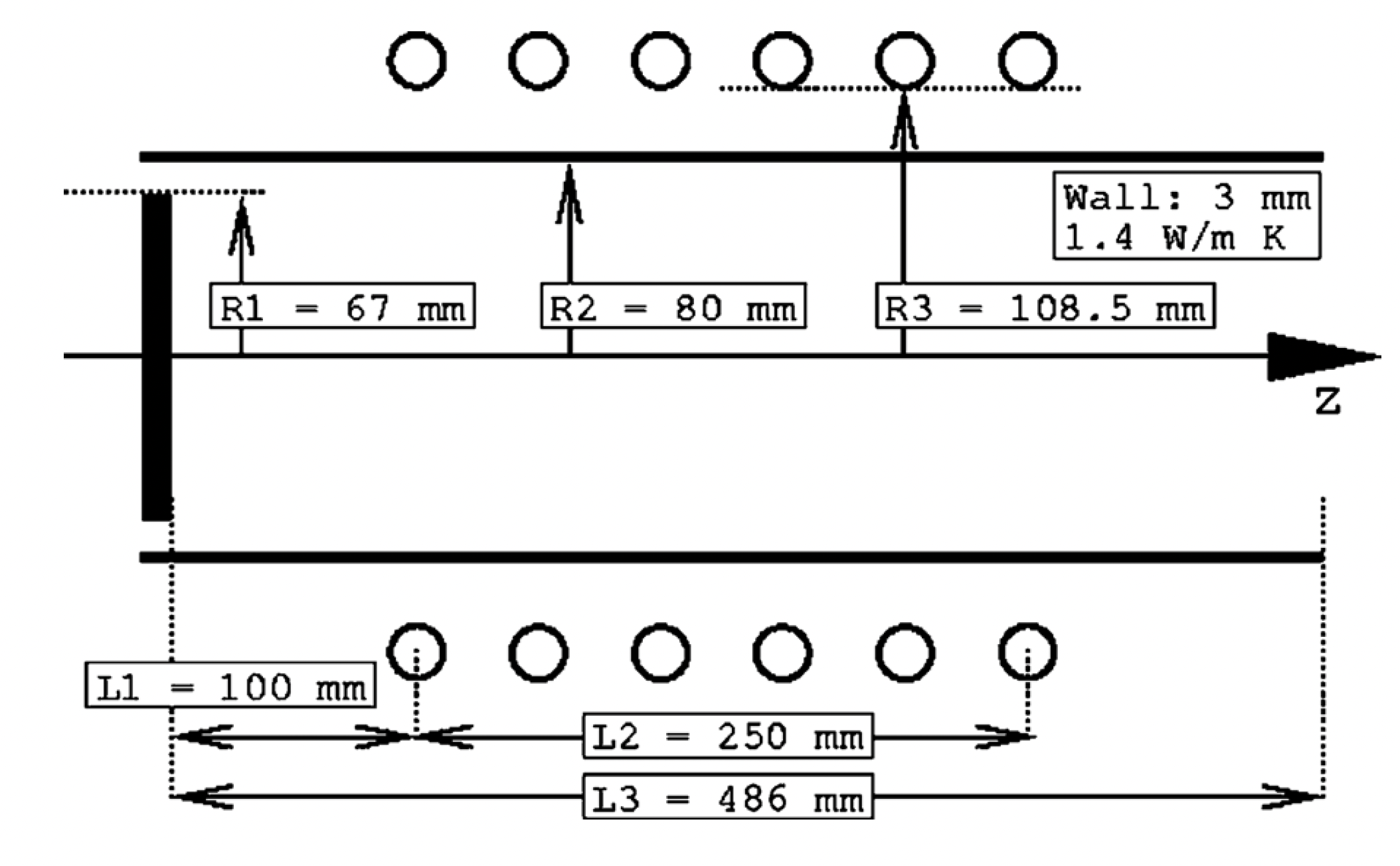}}
\caption{(a) Schematic of the ICP torch used for simulations (credits: Von Karman Institute for Fluid Dynamics\cite{abeele2000efficient}).}
\label{fig:torch}
\end{figure}

    2D axi-symmetric simulations of the ICP torch have been performed in this work for the same torch geometry as used in Paper I and is again shown in \cref{fig:torch}. Description of the grid, boundary conditions, and ambient conditions remain the same as used in Paper I. The frequency of the coils is again fixed to be \SI{0.45}{MHz} for all the simulations presented in this paper. The numerical framework for ICP simulations used in this work remains the same as in Paper I.
    \\

    \subsection{Plasma bubble morphology inside the ICP torch: vibronic StS versus 2T models}\label{sec:vibronic_vs_2T}
    This section presents a comparison of the plasma flow fields obtained from the vibronic StS simulations against those obtained from the conventional 2-T NLTE models. In a conventional 2-T NLTE model, each internal energy mode for all species follows a Maxwell-Boltzmann distribution at a specific temperature (translational and rotational modes at a single temperature T\textsubscript{h}, while free-electron, electronic and vibrational modes at a single temperature T\textsubscript{ev}). In vibronic StS simulations as discussed in Paper I, the electronic and vibrational states are treated as separate pseudo-species whose populations are determined by solving the vibronic Master equations fully coupled with the flow equations. The rotational modes are assumed to be in equilibrium with the translational modes at T\textsubscript{h}.

    \begin{figure}[!htb]
    \hspace*{-0.75cm}
    \subfloat[]{\includegraphics[scale=0.16]{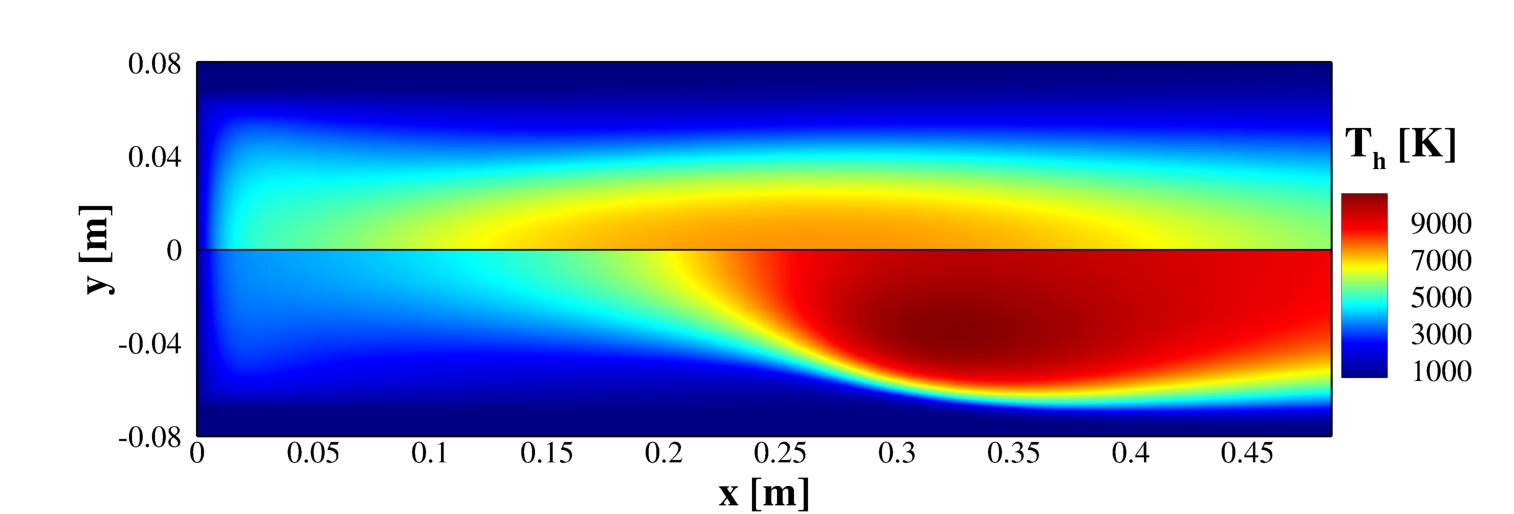}}
    \subfloat[]{\includegraphics[scale=0.16]{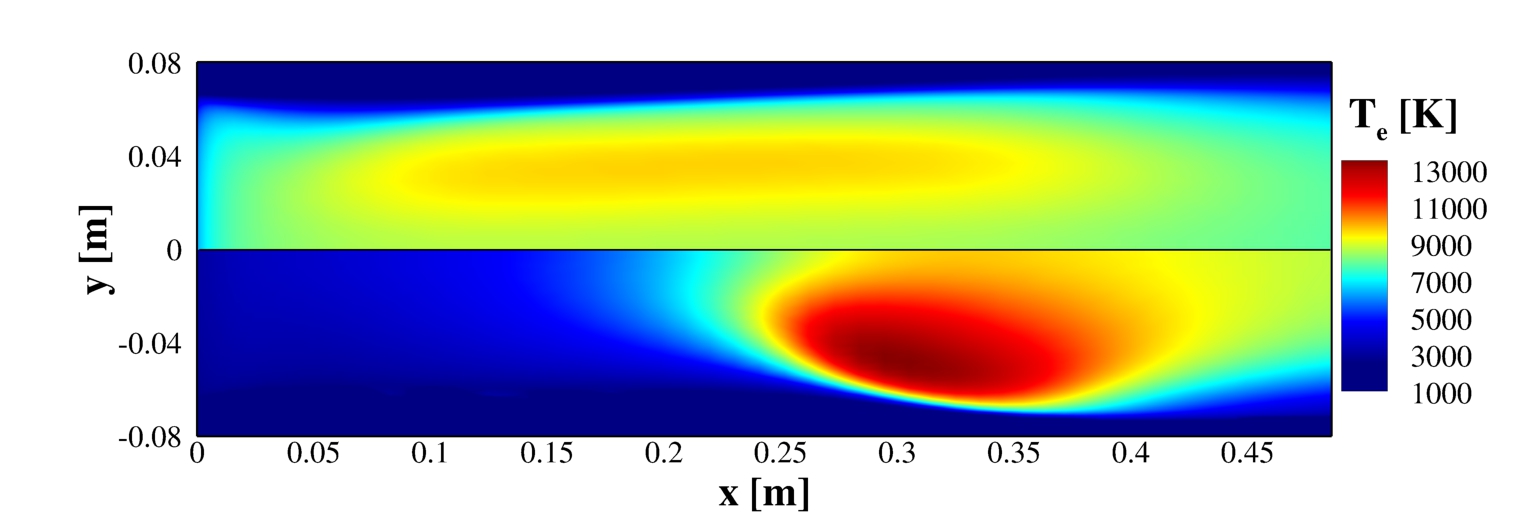}}
    \\
    \hspace*{-0.75cm}
    \subfloat[]{\includegraphics[scale=0.16]{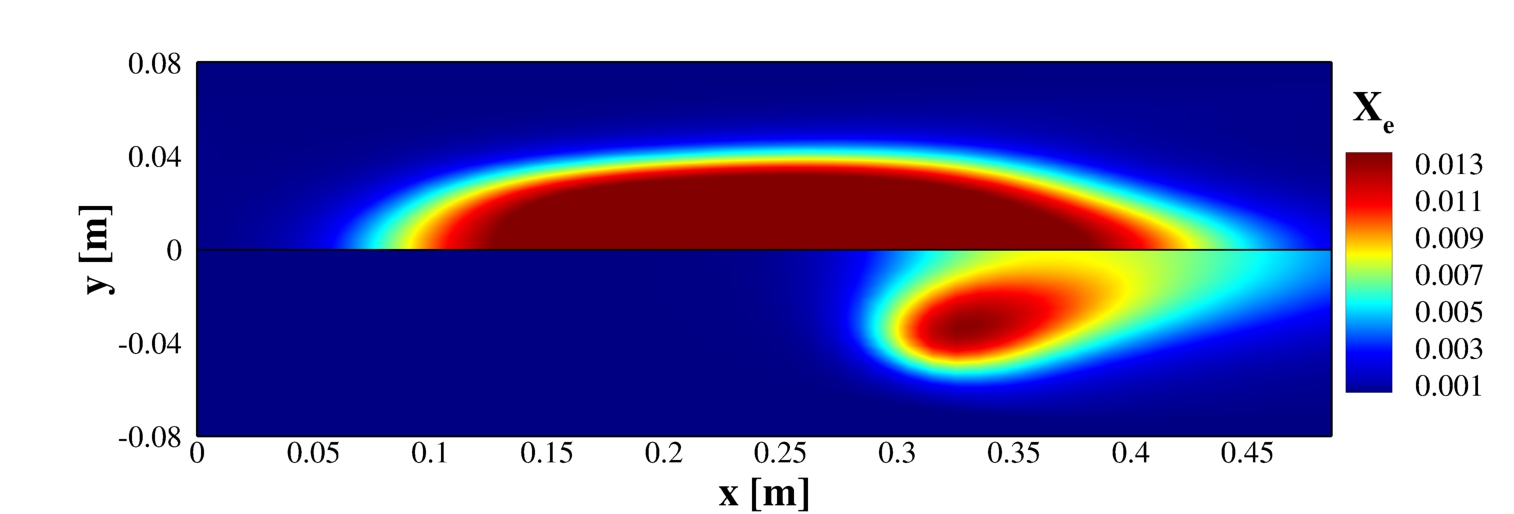}} \hspace{0.05in}
    \subfloat[]{\includegraphics[trim={0 3.75cm 0 3.75cm},clip, scale=0.26]{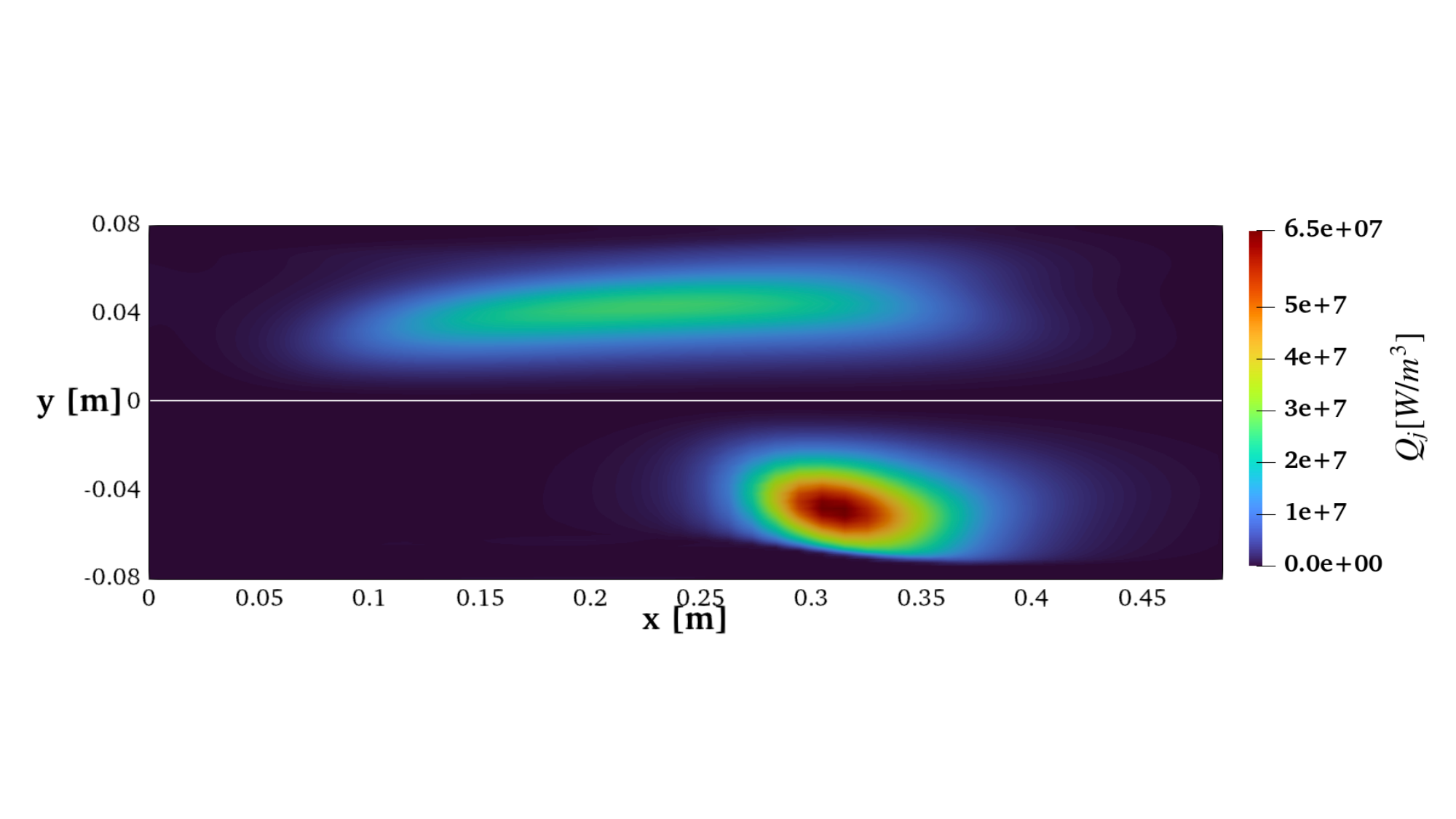}}
    \caption{Comparison of the plasma flow fields obtained from the vibronic StS and Park 2-T models: (a) heavy-species temperature, (b) electron temperature, (c) electron mole-fraction, and (d) Joule heating. Top: Park 2-T, Bottom: vibronic StS. Operating conditions: \SI{1000}{Pa}, \SI{50}{kW} and \SI{6}{g/s}.}
    \label{fig:vibronic_vs_park2T}
    \end{figure}

    \subsubsection{Vibronic StS versus Park 2-T }\label{sec:vibronic_vs_park2T}
    ICP torch Simulations were conducted under the following operating conditions: mass flow \SI{6}{g/s}, pressure \SI{1000}{Pa}, and power \SI{50}{kW}. The widely accepted Park 2-T model \cite{park2001chemical} was employed, which is the most widespread model for hypersonic applications. It is therefore instructive to compare the results from the new vibronic StS model against the conventional Park 2-T standard. In the case of the Park 2-T model, the default MW + Park's high-temperature correction formula is used for the vibrational relaxation times $\tau_s^{\mathrm{VT}}$ along with the non-preferential dissociation model for $\Omega^{\mathrm{CV}}$ term.

   \cref{fig:vibronic_vs_park2T} compares the plasma flow field obtained from the Park 2-T model with the one obtained from the vibronic StS model. It can be seen that the plasma flow field obtained from the two models is significantly different. In the Park 2-T simulation, Joule heating spans a more extensive volume, thus warming a larger plasma section. Given that the total power dissipated in the plasma is held constant, this results in a cooler overall plasma core. Contrarily, in the vibronic StS scenario, the electron concentration and thus the Joule heating is more localized within the coil region. This localization causes a downstream shift of the plasma core compared to the Park 2-T model's plasma core position. With heating concentrated in a smaller plasma section, the core's temperature is notably elevated due to the equivalent power dissipation in both cases. This substantial variability in plasma flow field confirms previous results from existing literature which suggests that ICP flow fields are highly sensitive to kinetic mechanisms \cite{zhang2016analysis}. Given that the reaction rates exhibit differences across distinct kinetic databases, the observed differences are to be expected. Moreover, inconsistent V-T relaxation times and the non-preferential dissociation model used in Park 2-T model could be another factor leading to this discrepancy. Hence, pinpointing the primary cause of this flow field deviation - be it the non-Boltzmann effect or differences in the kinetic database or parameters remains elusive.

     \begin{figure}[!htb]
    \hspace*{-0.75cm}
    \subfloat[]{\includegraphics[scale=0.16]{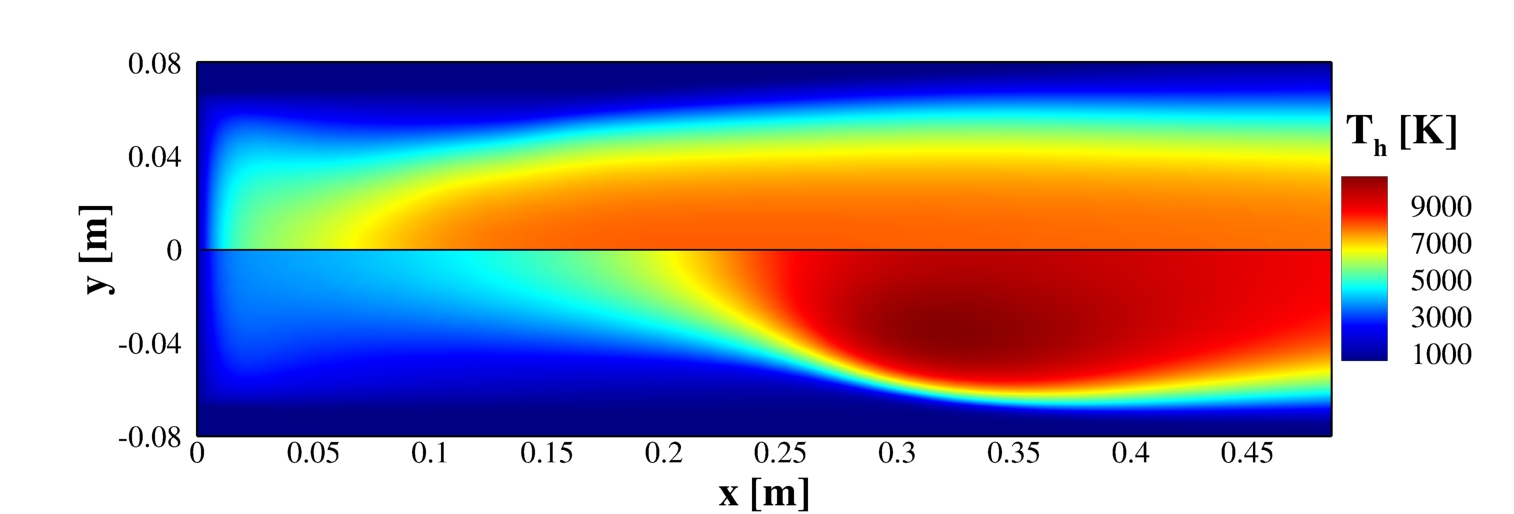}}
    \subfloat[]{\includegraphics[scale=0.16]{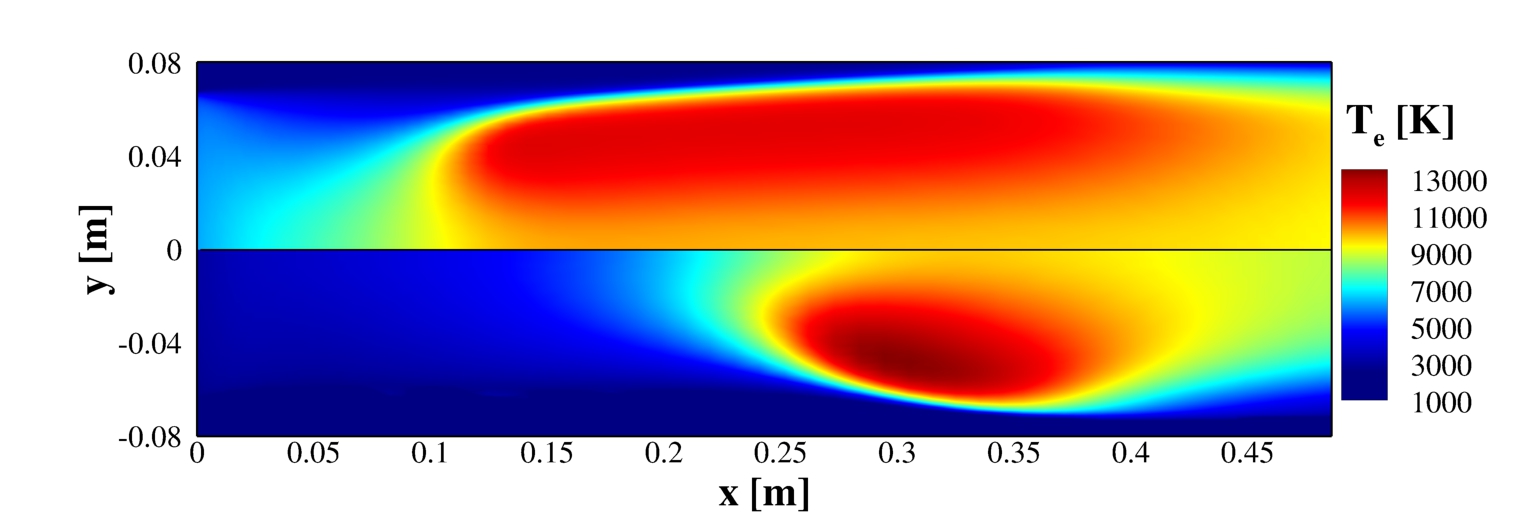}}
    \\
    \hspace*{-0.75cm}
    \subfloat[]{\includegraphics[scale=0.16]{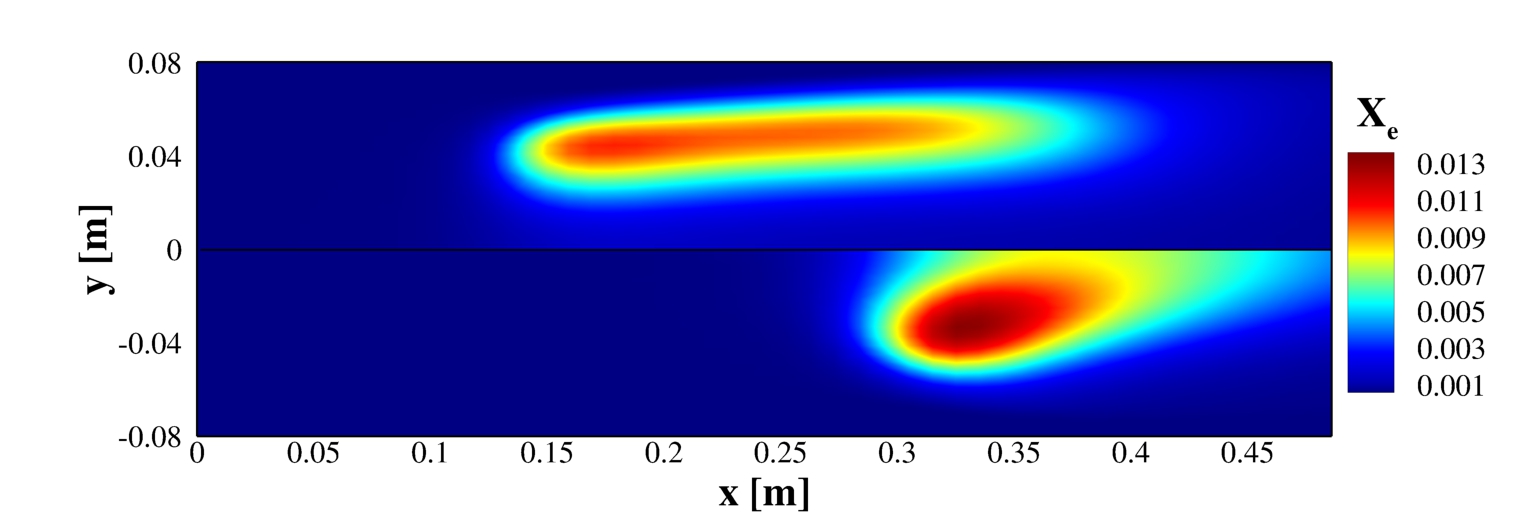}} \hspace{0.05in}
    \subfloat[]{\includegraphics[trim={0 3.75cm 0 3.75cm},clip,scale=0.26]{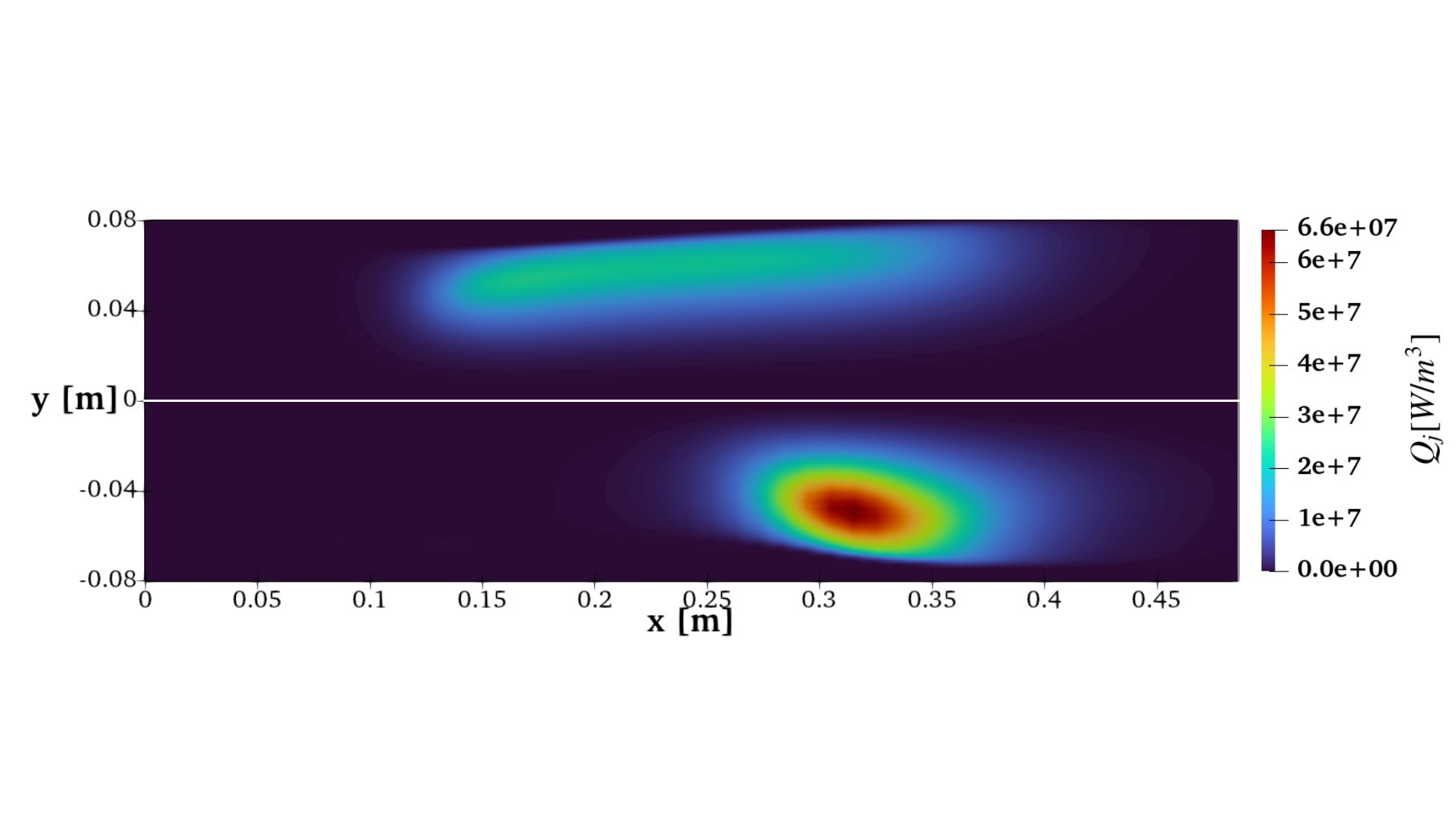}}
    \caption{Comparison of the plasma flow fields obtained from the vibronic StS and the consistent 2-T model (with inconsistent $\tau_{VT}$ and $\psi_s^D$): (a) heavy-species temperature, (b) electron temperature, (c) electron mole-fraction and (d) Joule heating. Top: 2-T, Bottom: vibronic StS. Operating conditions: \SI{1000}{Pa}, \SI{50}{kW} and \SI{6}{g/s}.}
    \label{fig:vibronic_vs_elio2T_inconsitent}
    \end{figure}

    \begin{figure}[!htb]
    \hspace*{-0.75cm}
    \subfloat[]{\includegraphics[scale=0.16]{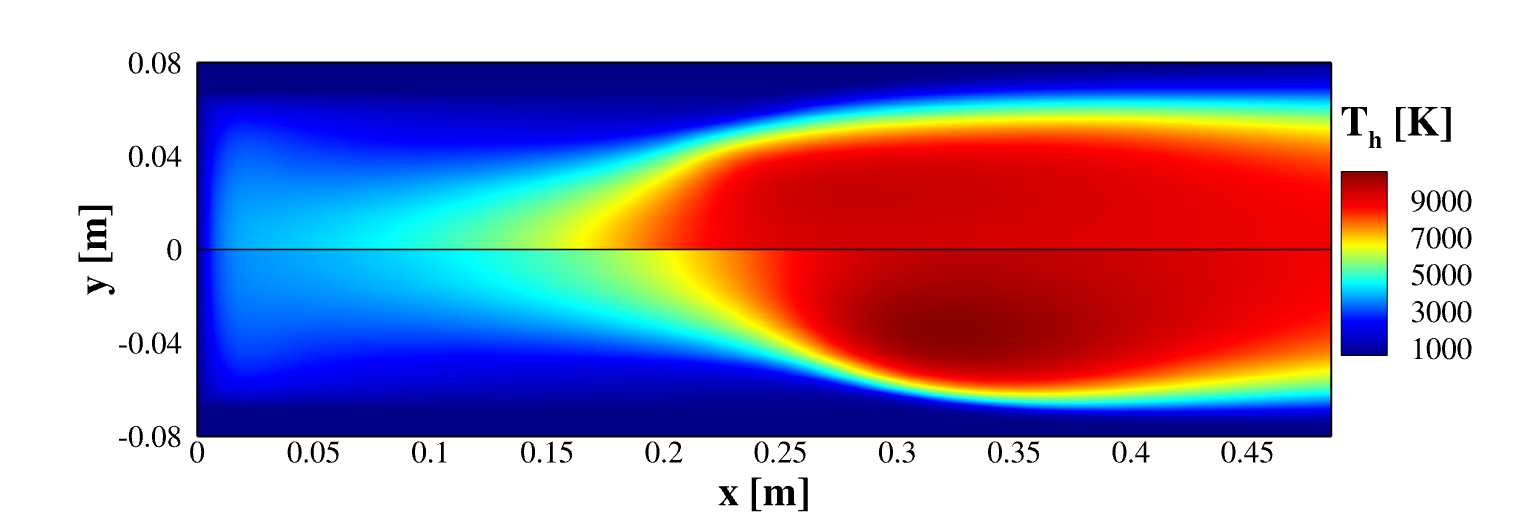}}
    \subfloat[]{\includegraphics[scale=0.16]{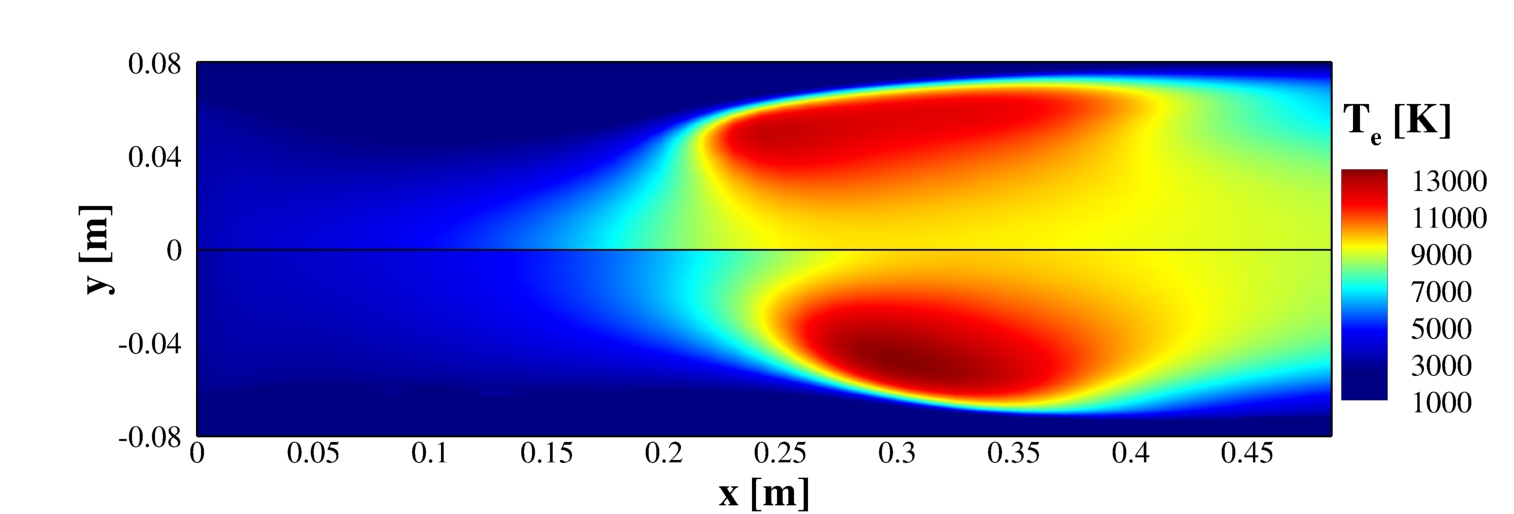}}
    \\
    \hspace*{-0.75cm}
    \subfloat[]{\includegraphics[scale=0.16]{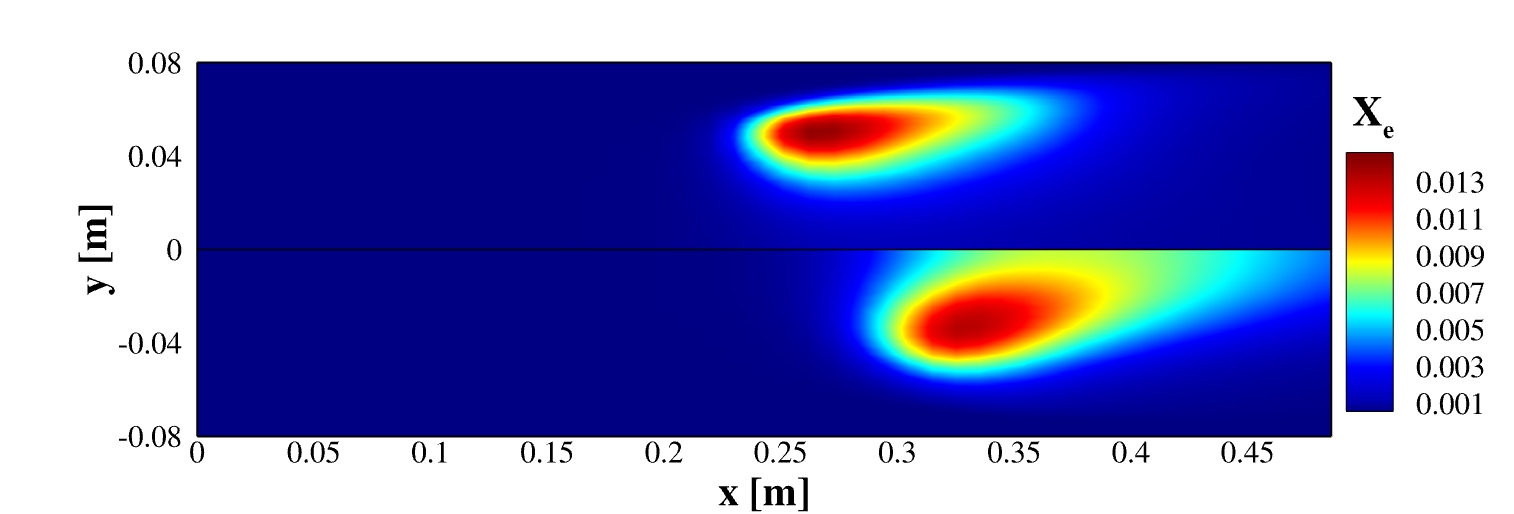}} \hspace{0.05in}
    \subfloat[]{\includegraphics[trim={0 3.75cm 0 3.75cm},clip,scale=0.26]{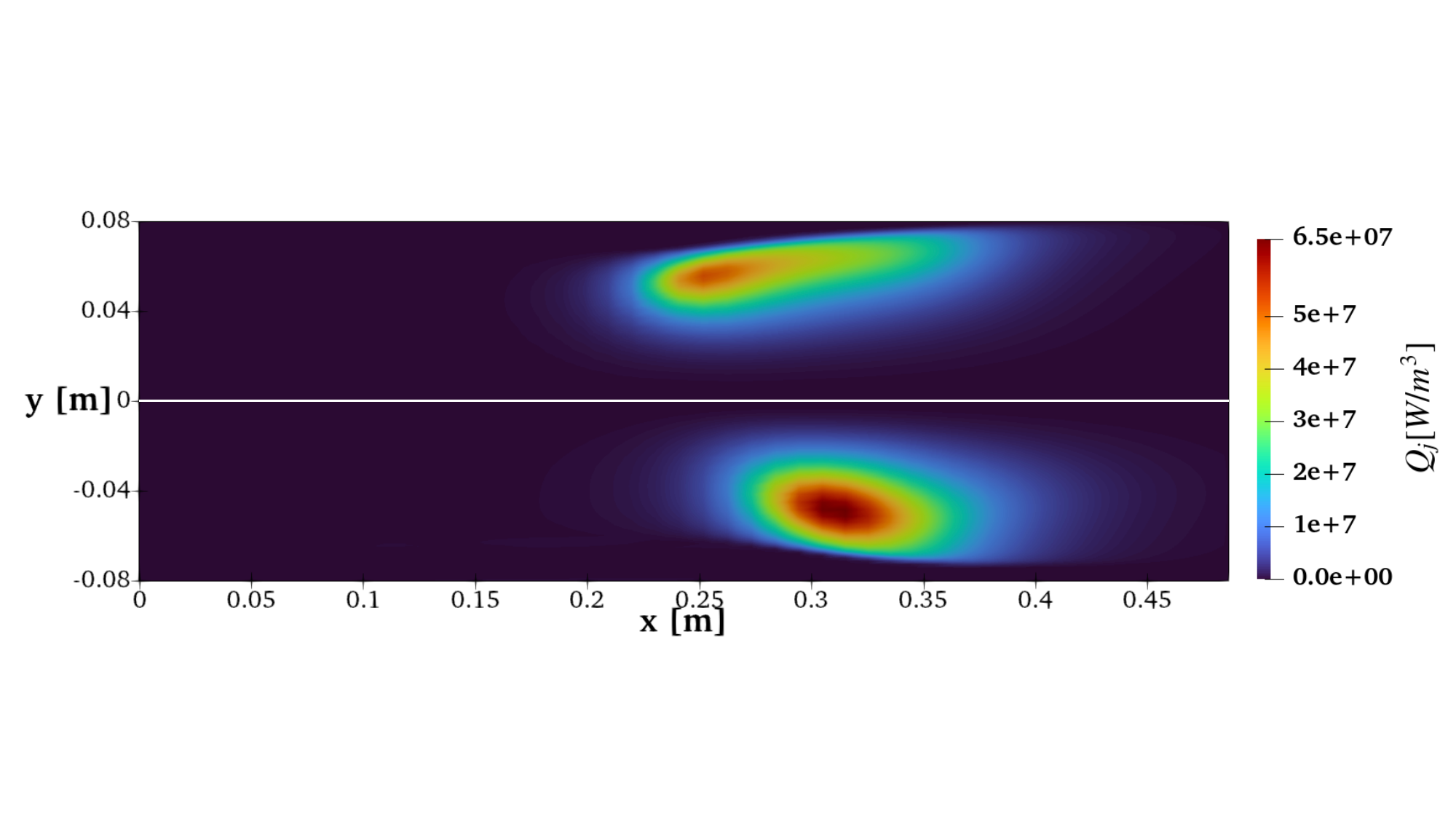}}
    \caption{Comparison of the plasma flow fields obtained from the vibronic StS and the consistent 2-T model (with consistent $\tau_{VT}$ but inconsistent $\psi_s^D$): (a) heavy-species temperature, (b) electron temperature, (c) electron mole-fraction and (d) Joule heating. Top: 2-T, Bottom: vibronic StS. Operating conditions: \SI{1000}{Pa}, \SI{50}{kW} and \SI{6}{g/s}.}
    \label{fig:vibronic_vs_elio2T_consistent_VT_only}
    \end{figure}

    \begin{figure}[!htb]
    \hspace*{-0.75cm}
    \subfloat[]{\includegraphics[scale=0.16]{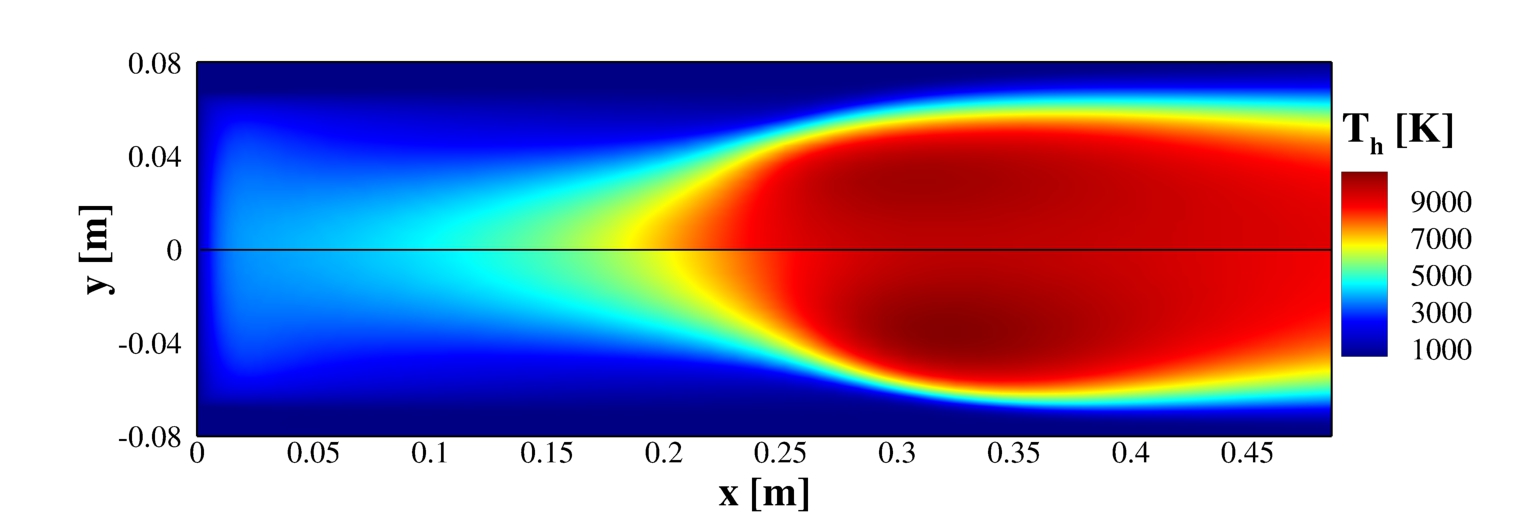}}
    \subfloat[]{\includegraphics[scale=0.16]{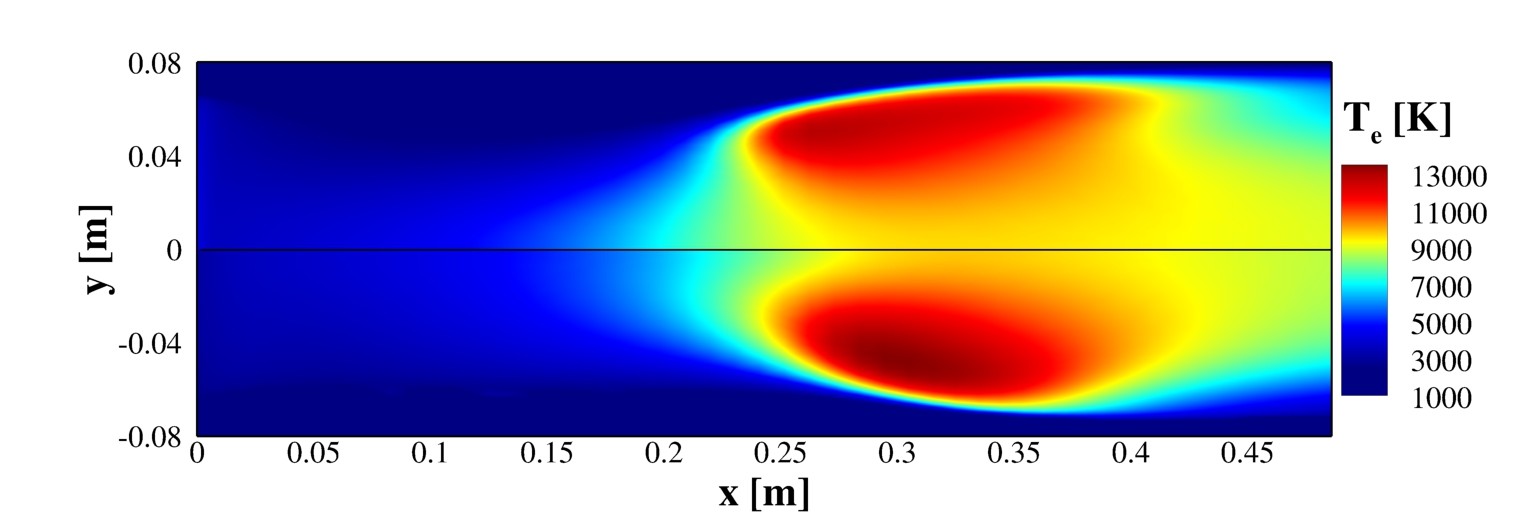}}
    \\
    \hspace*{-0.75cm}
    \subfloat[]{\includegraphics[scale=0.16]{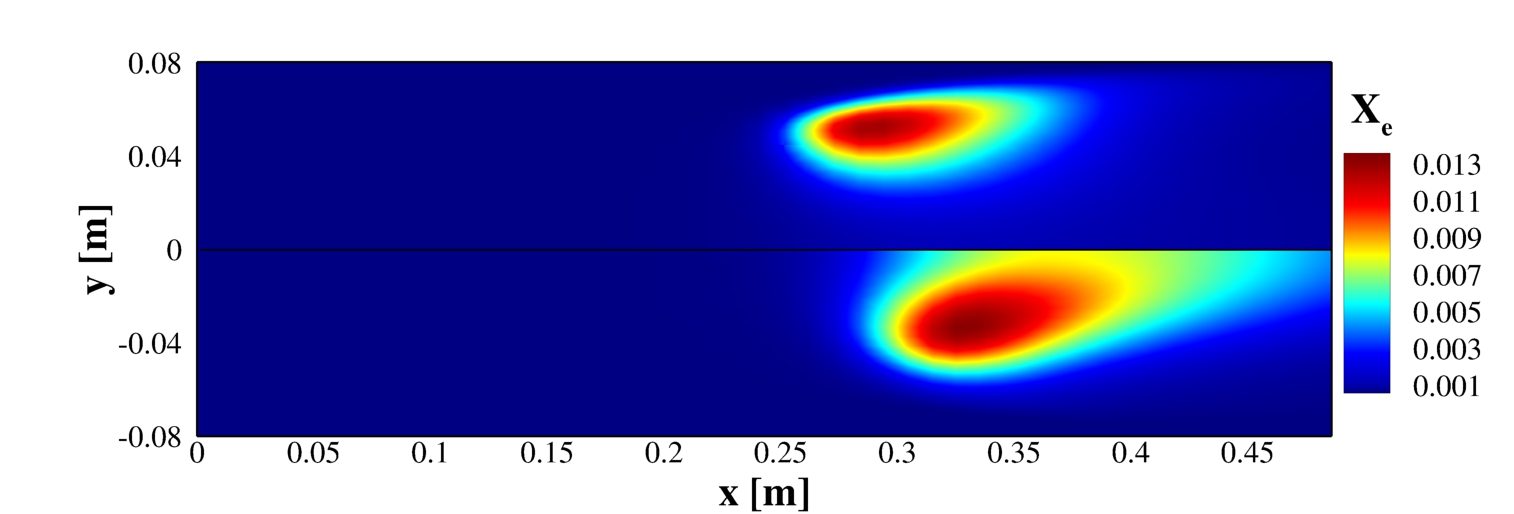}} \hspace{0.05in}
    \subfloat[]{\includegraphics[trim={0 3.75cm 0 3.75cm},clip,scale=0.26]{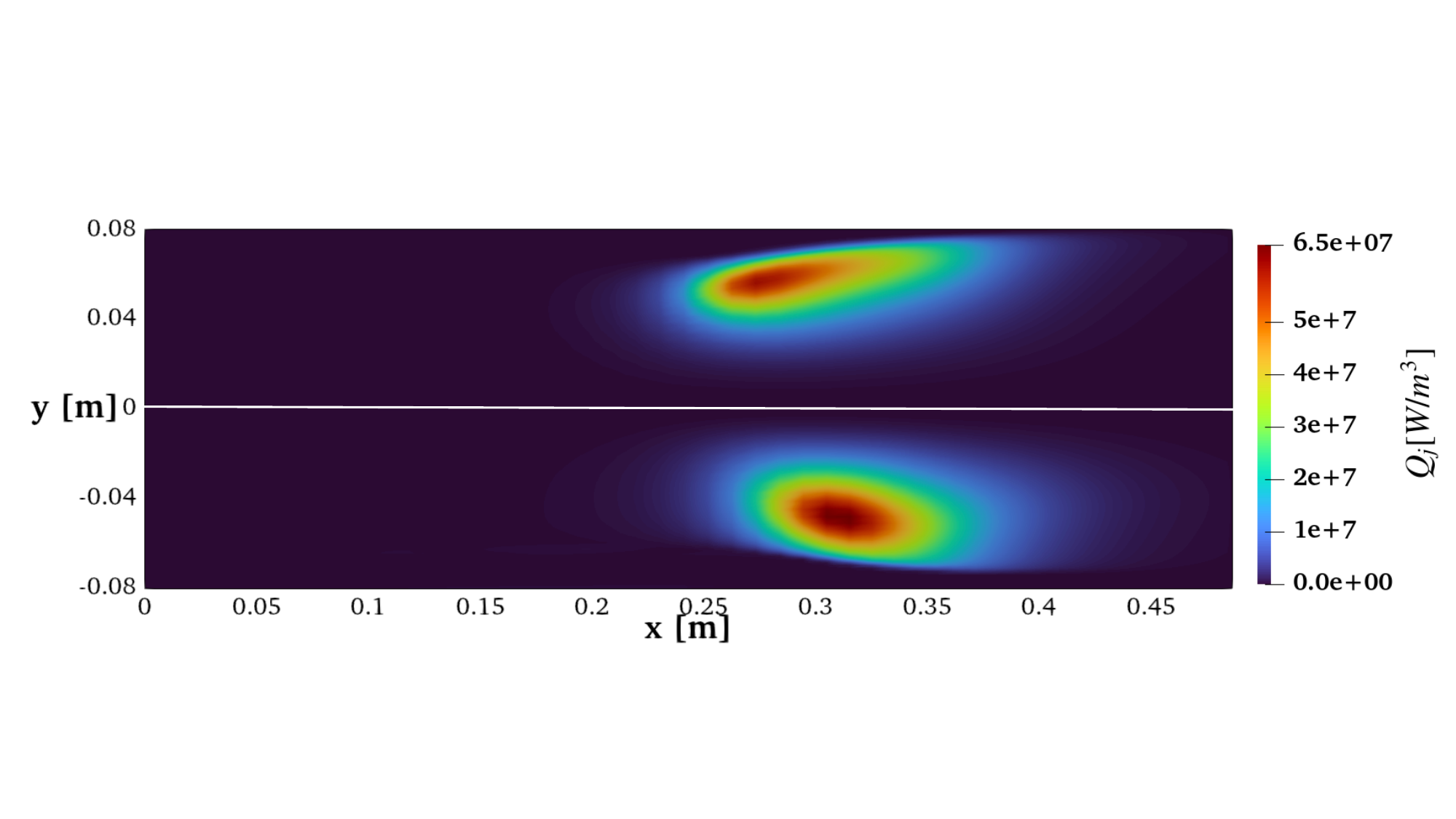}}
    \caption{Comparison of plasma the flow fields obtained from the vibronic StS and the consistent 2-T model (with consistent $\tau_{VT}$ and $\psi_s^D$): (a) heavy-species temperature, (b) electron temperature, (c) electron mole-fraction and (d) Joule heating. Top: 2-T, Bottom: vibronic StS. Operating conditions: \SI{1000}{Pa}, \SI{50}{kW} and \SI{6}{g/s}.}
    \label{fig:vibronic_vs_elio2T_consistent_VT_CV}
    \end{figure}

    \subsubsection{Vibronic StS versus consistent 2-T}
    Subsequent simulations were carried out employing the consistent 2-T model, derived from the vibronic StS model, as detailed in \cref{sec:reduction}. Initially, only the consistent chemical reaction rates were incorporated into the 2-T simulation. The $\tau_s^{\mathrm{VT}}$ and $\Omega^{\mathrm{CV}}$ terms, however, were retained as per the Park 2-T model. This approach was chosen to specifically examine the influence of modified reaction rates on the plasma flow field. \cref{fig:vibronic_vs_elio2T_inconsitent} offers a comparative analysis that contrasts the plasma flow field resulting from the consistent 2-T model with that derived from the vibronic StS simulation. Even though the flow field from the consistent 2-T model diverges slightly from the Park 2-T flow field due to the updated reaction rate parameters, it remains significantly different from the vibronic StS flow field. The qualitative flow characteristics yielded by the consistent 2-T model align more closely with those from the Park 2-T model than with the vibronic StS model. The macroscopic rates for pivotal reactions in ICP (such as heavy-impact dissociation and electron-impact ionization\cite{zhang2016analysis}) obtained by refining the vibronic StS kinetics are remarkably akin to Park’s macroscopic rates (for details, refer to \cref{appendix:park_vs_elio_rates}). This underlines that the pronounced disparities in plasma core morphology between the vibronic StS and the 2-T simulations are not predominantly ascribed to variations in reaction rates.

    Subsequently, consistent $\tau_s^{\mathrm{VT}}$ expressions, as outlined in \cref{tab:tauVT}, were integrated into the 2-T model. This was done instead of the default MW plus high-temperature correction expressions. Meanwhile, the $\Omega^{\mathrm{CV}}$ term remained modeled by the non-preferential dissociation model. \cref{fig:vibronic_vs_elio2T_consistent_VT_only} offers a comparative view of the plasma flow field within the torch, showcasing the results from the reduced 2-T model (with consistent $\tau_s^{\mathrm{VT}}$) against those from the vibronic StS model. Interestingly, by merely adjusting the $\tau_s^{\mathrm{VT}}$ parameter, the plasma core is shifted downstream, making the plasma flow field more reminiscent of the vibronic StS flow field. This pivotal observation underscores that the plasma core's position can be modulated by a single term - the vibrational relaxation time. This further highlights the crucial role of modeling this term, drawing on accurate state-to-state kinetic calculations.
    
    Next, the preferential dissociation model for $\Omega^{\mathrm{CV}}$ delineated in \cref{sec:reduction} is integrated into the 2-T model. For this paper, the 2-T model, encompassing consistent rates derived from the vibronic StS model coupled with the consistent $\tau_s^{\mathrm{VT}}$ and $\Omega^{\mathrm{CV}}$ terms, will be denoted as the \textquotedblleft consistent 2-T model\textquotedblright. \cref{fig:vibronic_vs_elio2T_consistent_VT_CV} compares the plasma flow field derived from the consistent 2-T model, with that from the vibronic StS model. The consistent 2-T model is observed to produce a flow field that aligns qualitatively with the vibronic StS model, especially in terms of plasma core location, shape, and peak temperature. This demonstrates that the consistent 2-T model can efficiently capture the non-equilibrium effects computed by the StS model but with a significantly reduced computational burden.
    That said, as shown in \cref{fig:vibronic_vs_elio2T_consistent_VT_CV} (c), there remains a pronounced discrepancy in electron concentration between the 2-T and vibronic StS simulations. This disparity stems from the non-QSS effect as well as the limitations of the present 2-T model in accounting for phenomena like non-preferential ionization, electron-impact electronic excitation, dissociation, and others. While including consistent energy transfer terms into the energy equation of the 2-T model allows one to obtain agreement with the StS model for the temperature fields, it's important to note that a 2-T model cannot accurately capture the non-Boltzmann distribution of internal states. Hence, for operating conditions where there might be a significant non-Boltzmann and non-QSS effect inside the ICP facility, a fully coupled state-to-state approach is still needed if accurate prediction of internal state populations is required (such as for reconstructing the spectra obtained from optical emission spectroscopy data from experiments, radiation coupling, \emph{etc.}).

\begin{figure}[!htb]
    \hspace*{-0.75cm}
    \subfloat[][]{\includegraphics[scale=0.55]{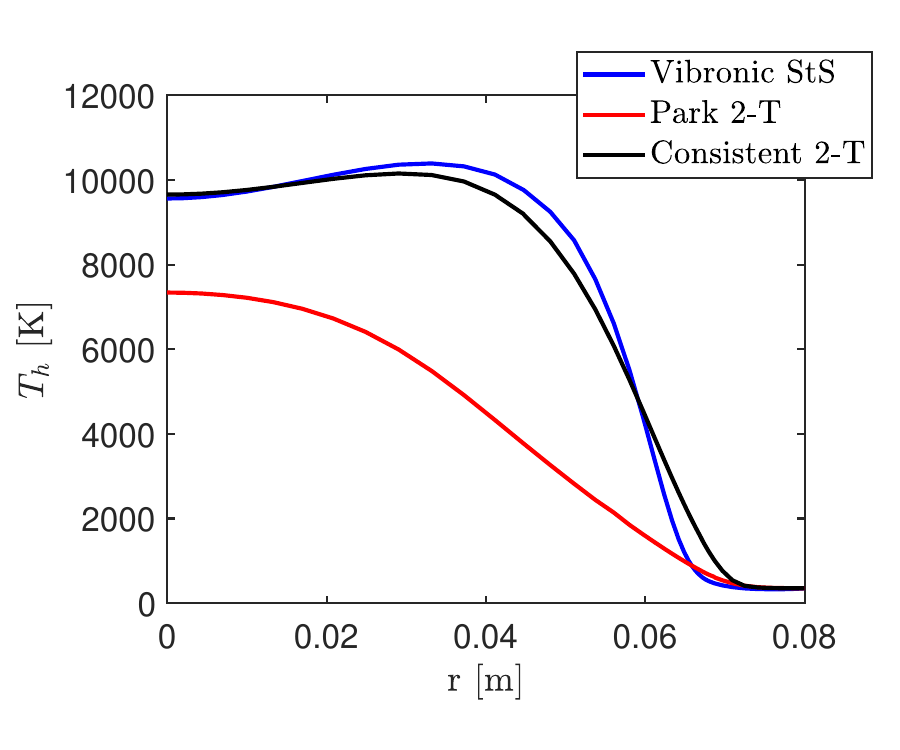}}
    \subfloat[][]{\includegraphics[scale=0.55]{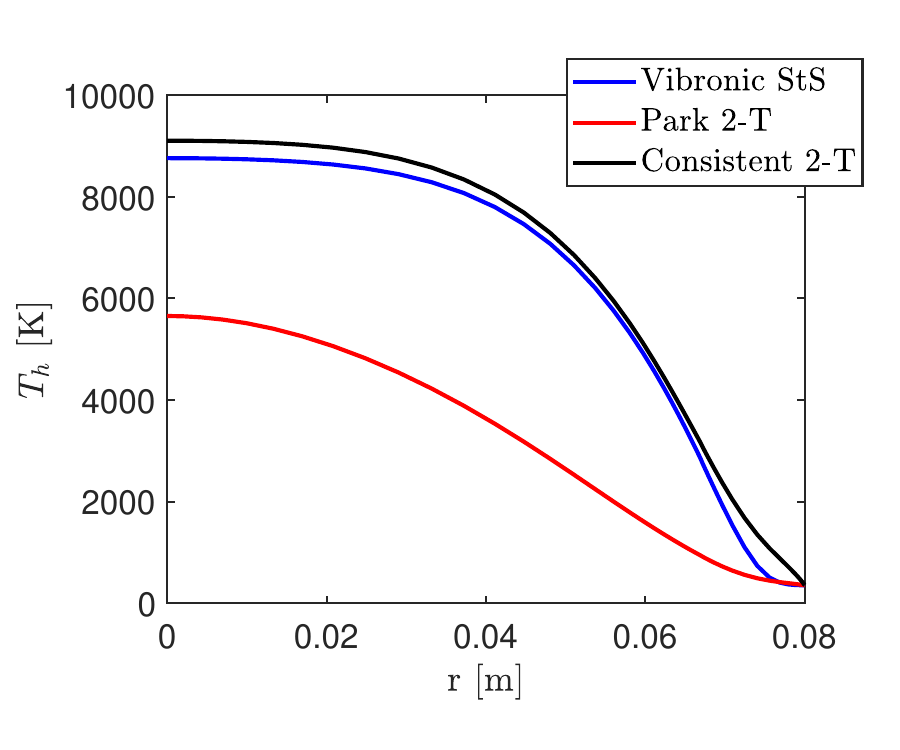}}
    \caption{Heavy-species temperature profiles at: (a) x = \SI{0.3}{m} (mid-torch location) and (b) x = \SI{0.485}{m} (torch outlet). Operating conditions: \SI{1000}{Pa}, \SI{50}{kW} and \SI{6}{g/s}.}
    \label{fig:vib_park_elio_Th}
    \end{figure}

\begin{figure}[!htb]
    \hspace*{-0.75cm}
    \subfloat[][]{\includegraphics[scale=0.55]{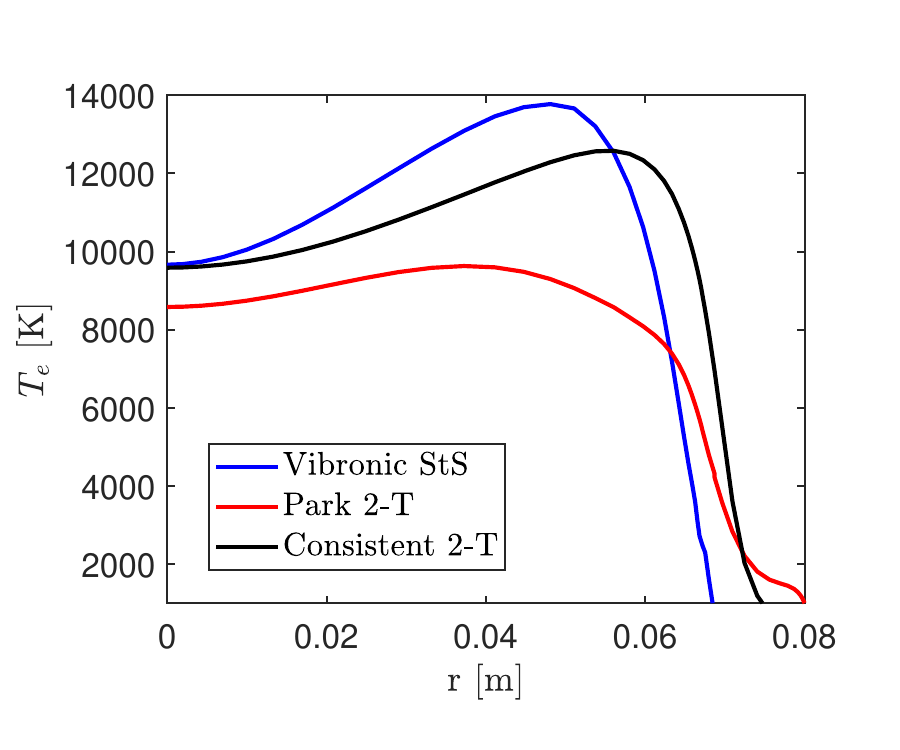}}
    \subfloat[][]{\includegraphics[scale=0.55]{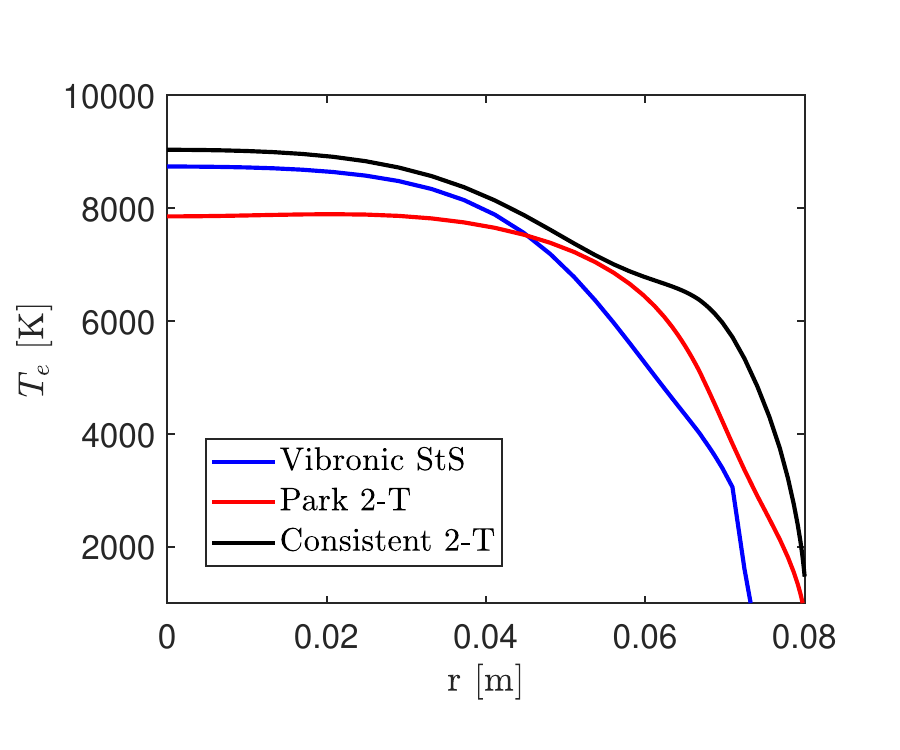}}
    \caption{Electro-vibrational/free-electron temperature profiles at: (a) x = \SI{0.3}{m} (mid-torch location) and (b) x = \SI{0.485}{m} (torch outlet). Operating conditions: \SI{1000}{Pa}, \SI{50}{kW} and \SI{6}{g/s}.}
    \label{fig:vib_park_elio_Te}
    \end{figure}

\begin{figure}[!htb]
    \hspace*{-0.75cm}
    \subfloat[][]{\includegraphics[scale=0.55]{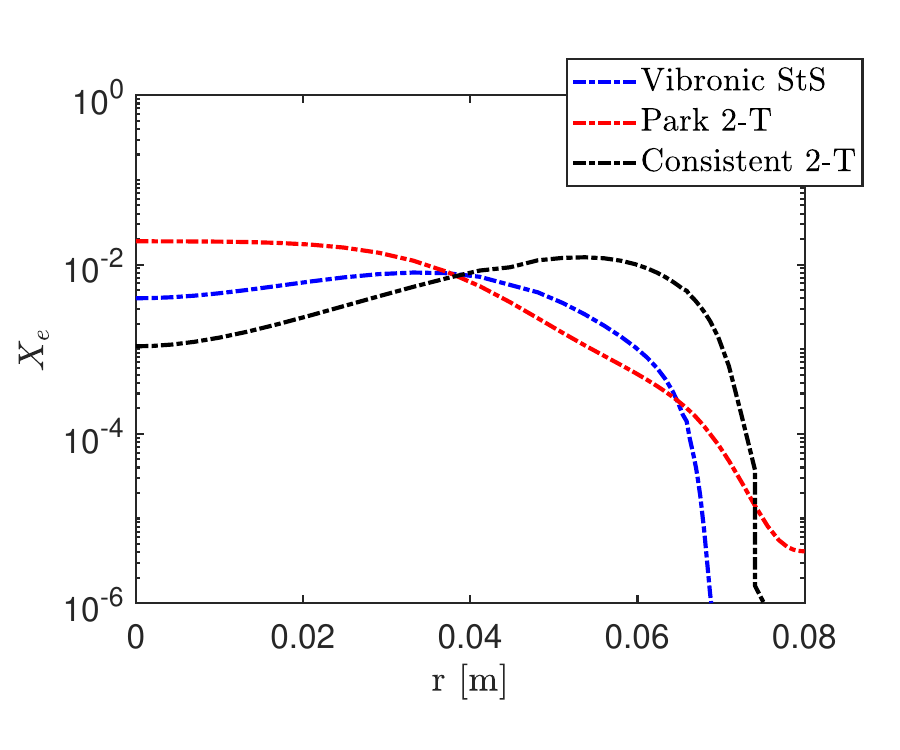}}
    \subfloat[][]{\includegraphics[scale=0.55]{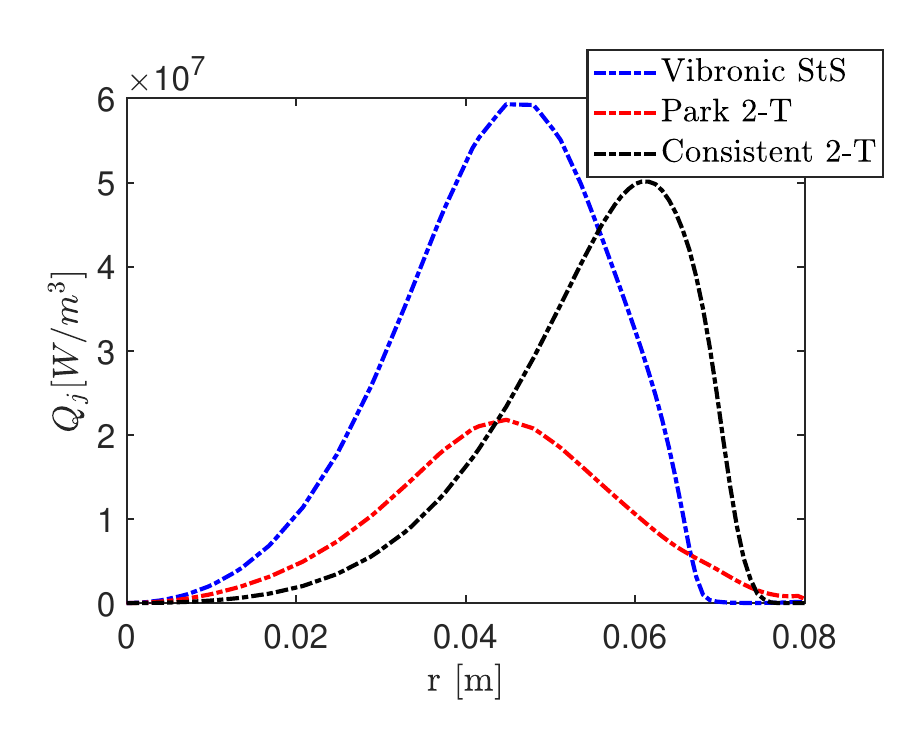}}
    \caption{(a) Electron mole-fraction and (b) Joule heating profiles at x = \SI{0.3}{m} (mid-torch location). Operating conditions: \SI{1000}{Pa}, \SI{50}{kW} and \SI{6}{g/s}.}
    \label{fig:vib_park_elio_Xe_Qj}
    \end{figure}

    \cref{fig:vib_park_elio_Th} and \cref{fig:vib_park_elio_Te} show the plasma temperature profiles obtained from various models (i.e. vibronic StS, Park 2-T, and consistent 2-T model) re-affirming that the consistent 2-T model is able to reproduce the vibronic StS gas temperature profiles, while there are small differences in the electron temperature profiles. \cref{fig:vib_park_elio_Xe_Qj} (a) shows the radial electron concentration profiles at the mid-torch location depicting that while neither Park nor the consistent 2-T model matches the concentration profiles exactly, the consistent 2-T model still gives much better agreement with the StS profiles. \cref{fig:vib_park_elio_Xe_Qj} (b) shows the Joule heating profiles at the mid-torch location showing significantly lower peak value in case of Park 2-T which is a result of distribution of the Joule heating over a larger plasma volume, unlike the case of vibronic StS and the consistent 2-T where the Joule heating distribution is localized to a smaller region and hence has much larger peak values to have the same dissipated power. 
    
    Finally, this section further suggests that the heavy-impact vibrational excitation and dissociation of N\textsubscript{2} has the first-order effect on the plasma core in terms of its location, shape and peak temperatures, as these kinetics are the ones used in modeling the $\Omega^{VT}$ and $\Omega^{CV}$ energy transfer terms in the internal energy equation of the 2-T model.

    \subsection{Comparative study of the plasma flow fields obtained from various physico-chemical models}

    This subsection presents a comparative study of the plasma flow field inside the ICP torch obtained using different physico-chemical models. To test the ability of the vibronic StS model to provide physically consistent results for a wide range of operating conditions, simulations were conducted for two extreme operating conditions: a high-pressure case (\SI{10000}{Pa}, \SI{50}{kW}) and a high power case (\SI{1000}{Pa}, \SI{250}{kW}) apart from the low pressure and low power base case (\SI{1000}{Pa}, \SI{50}{kW}) already discussed in the previous subsection. The mass flow (\SI{6}{g/s}) remains the same as before for all the simulations. The internal temperatures of various components denoted as $\mathrm{T}_{\mathrm{N}_2}$, $\mathrm{T}_{\mathrm{N}_2^+}$, $\mathrm{T}_\mathrm{N}$ and $\mathrm{T}_{\mathrm{N}^+}$ in the figures have been computed using the populations of vibronic (for molecules) and electronic (for atoms) states as defined in Paper I.  
     
    \subsubsection{Base case $\left(\SI{1000}{Pa}, \SI{50}{kW}\right)$}
\begin{figure}[!htb]
    \hspace*{-0.75cm}
    \subfloat[][]{\includegraphics[scale=0.16]{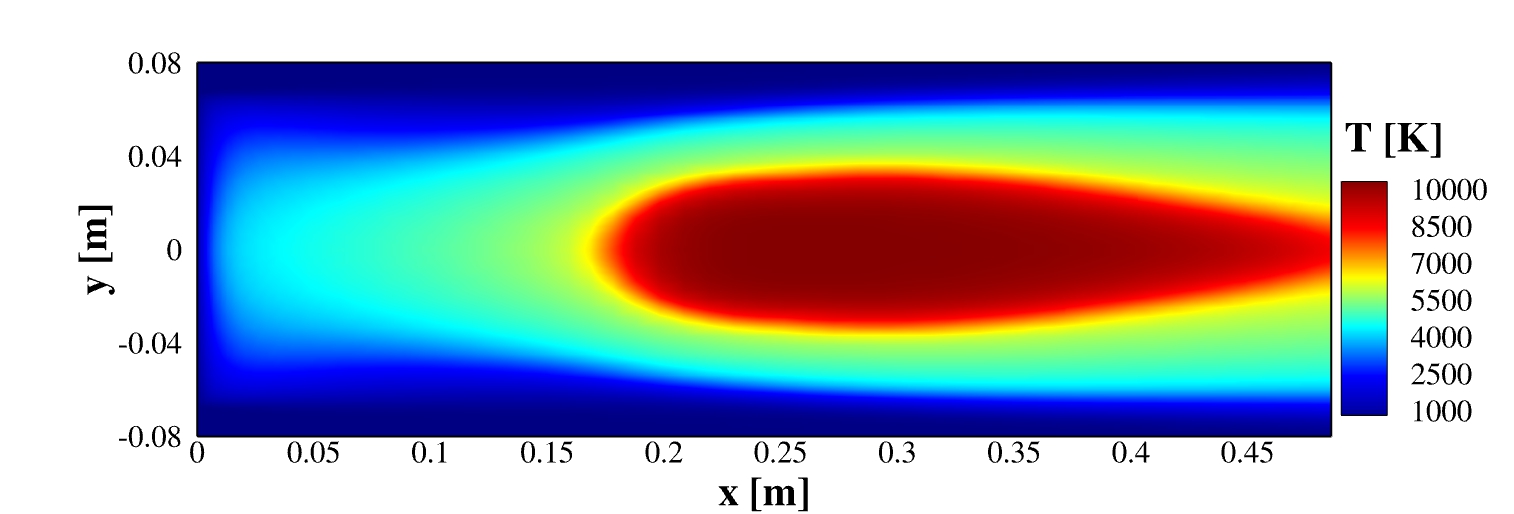}}
    \subfloat[][]{\includegraphics[scale=0.16]{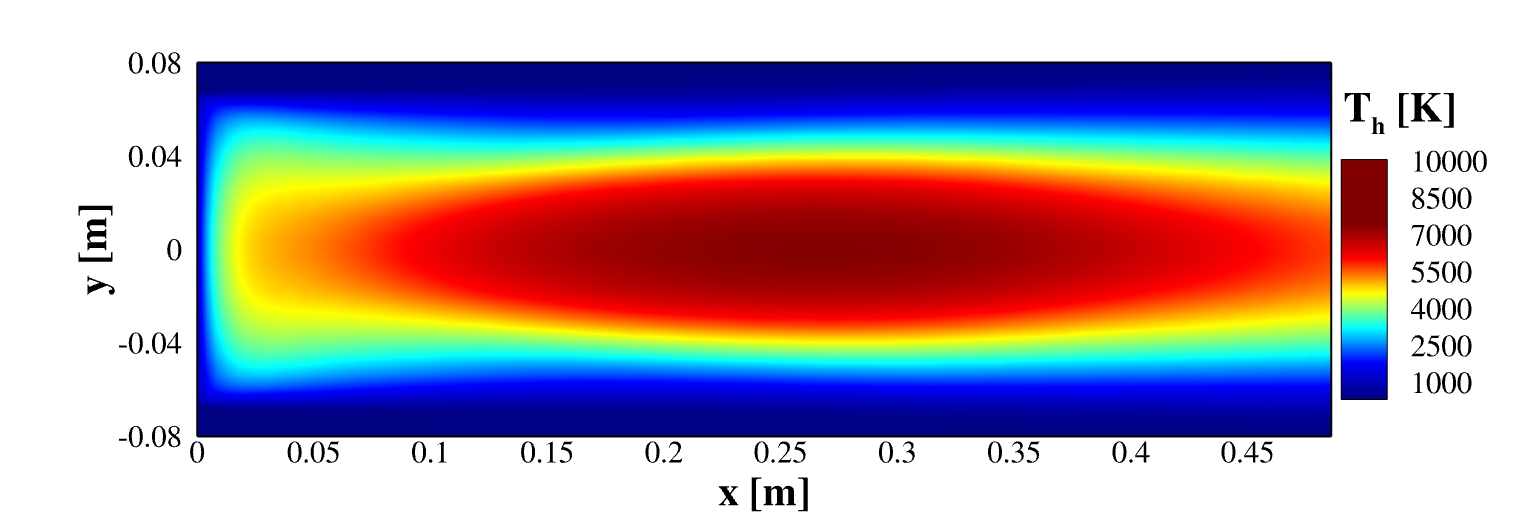}}\\
    \hspace*{-0.75cm}
    \subfloat[][]{\includegraphics[scale=0.16]{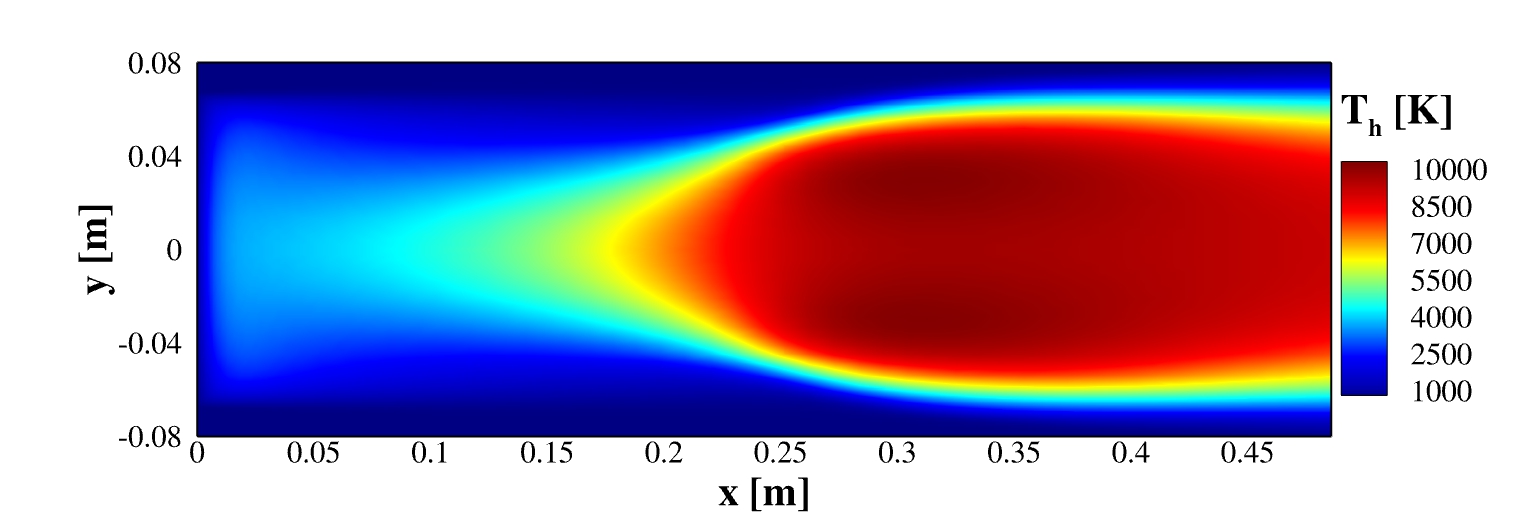}}
    \subfloat[][]{\includegraphics[scale=0.16]{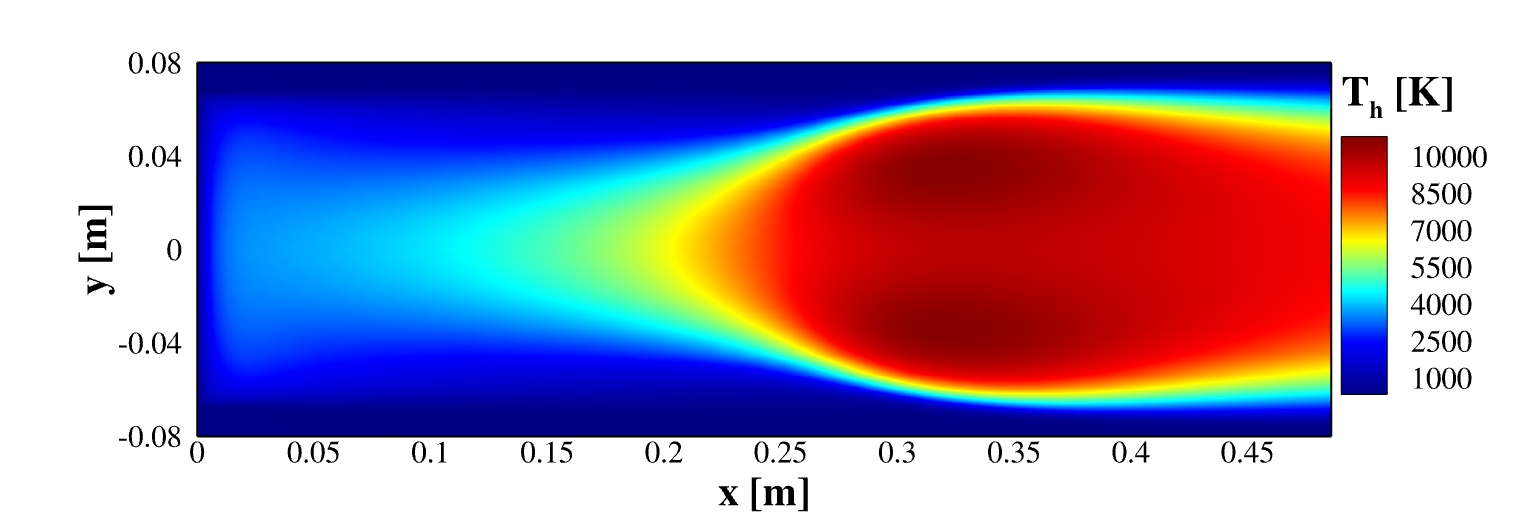}}
    \caption{Plasma heavy-species temperature contours: (a) LTE, (b) Park 2-T, (c) consistent 2-T, and (d) vibronic StS. Operating conditions: \SI{1000}{Pa}, \SI{50}{kW} and \SI{6}{g/s}.}
    \label{fig:T_contour_model_compare_1KPa_50KW}
    \end{figure}

\begin{figure}[!htb]
    \hspace*{-0.75cm}
    \subfloat[][]{\includegraphics[scale=0.55]{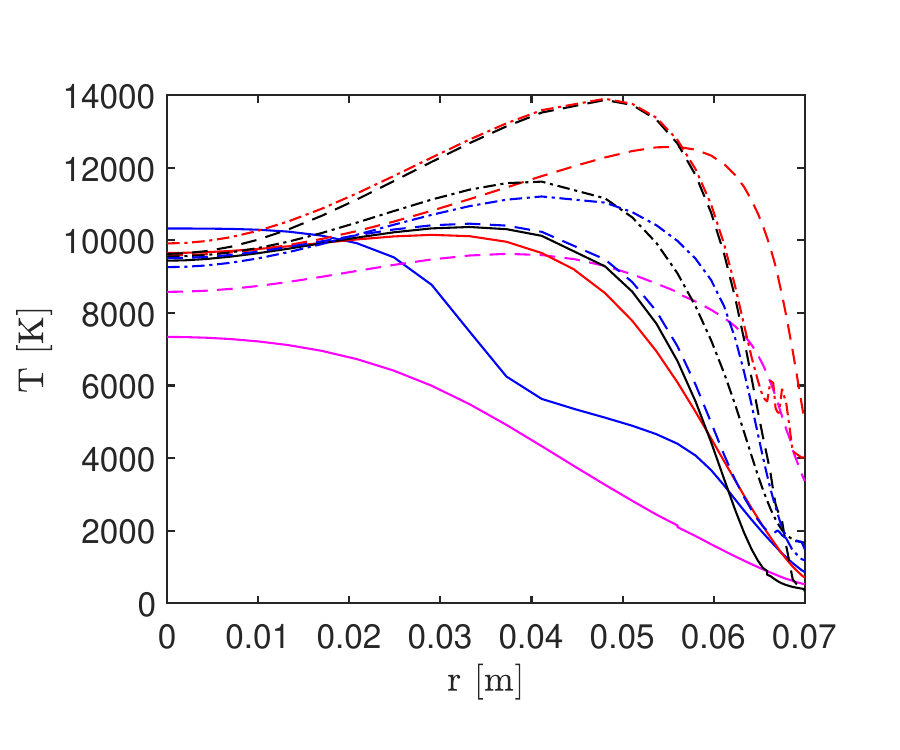}}
    \subfloat[][]{\includegraphics[scale=0.55]{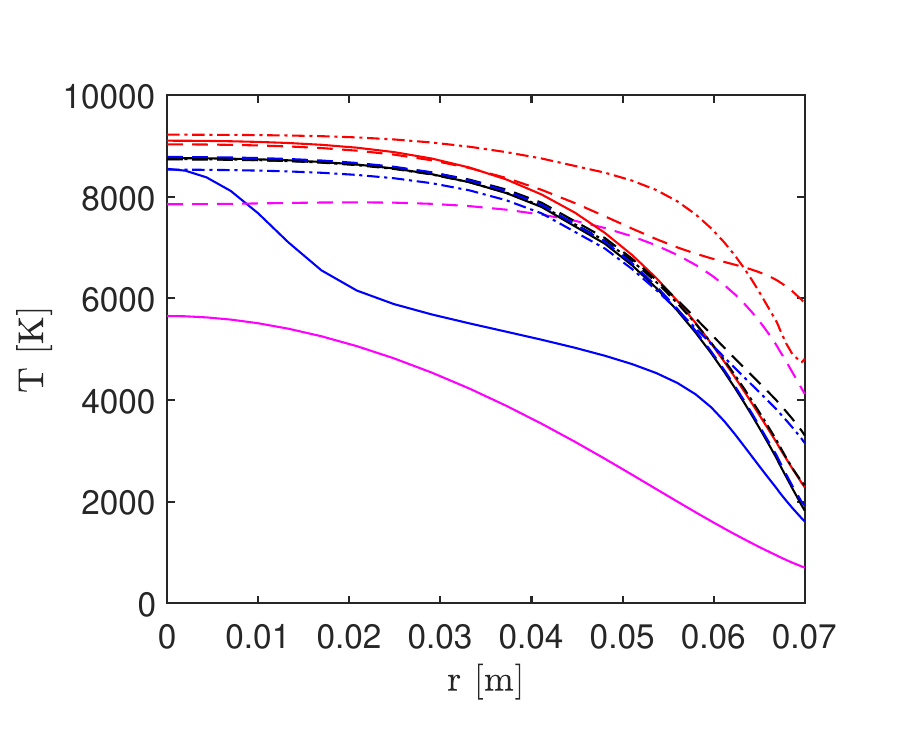}}
    \caption{Radial temperature profiles at: (a) x = \SI{0.3}{m} (mid-torch location) and (b) x = \SI{0.485}{m} (torch outlet). Solid blue line: T (LTE), solid magenta line: T\textsubscript{h} (Park 2-T), dashed magenta line: T\textsubscript{ev} (Park 2-T), solid red line: T\textsubscript{h} (consistent 2-T), dashed red line: T\textsubscript{ev} (consistent 2-T), solid black line: T\textsubscript{h} (vibronic StS), dashed black line: T\textsubscript{e} (vibronic StS), dashed-dot black line: T\textsubscript{N} (vibronic StS), dashed-dot blue line: T\textsubscript{N\textsubscript{2}} (vibronic StS), dashed blue line: T\textsubscript{N\textsuperscript{+}} (vibronic StS), and dashed-dot red line: $\mathrm{T}_{\mathrm{N}_2^+}$ (vibronic StS). Operating conditions: \SI{1000}{Pa}, \SI{50}{kW} and \SI{6}{g/s}.}
    \label{fig:T_profiles_model_compare_1KPa_50KW}
    \end{figure}

    This operating condition as discussed in previous sub-sections is characterized by highly non-equilibrium and non-Boltzmann conditions which necessitates the use of state-of-the-art NLTE models to accurately capture the NLTE effects. The vibronic StS model works quite well for this condition giving physically consistent plasma flow field as discussed in Paper I. However, for the sake of completeness, the base case flow fields have been presented for all the physico-chemical models in this subsection. \cref{fig:T_contour_model_compare_1KPa_50KW} shows the plasma temperature contours inside the ICP torch obtained from various physico-chemical models. The differences between Park 2-T and consistent 2-T flow fields have already been discussed in the previous subsection, while the comparison between LTE and vibronic StS flow fields was discussed in Paper I. \cref{fig:T_profiles_model_compare_1KPa_50KW} shows the radial temperature profiles inside the ICP torch obtained from various models. It is interesting to observe that at the torch outlet (\textit{i.e.} far from coils), park 2-T temperature profiles still show significant non-equilibrium between the translational and the electro-vibrational modes, whereas the vibronic StS and consistent 2-T models are close to thermal equilibrium between various modes. This observation agrees with the fact that the V-T relaxation time for $\mathrm{N}_2-\mathrm{N}$ system as shown in \cref{fig:tauVT} used in Park 2-T model is much higher than the one obtained from state-to-state kinetics and hence Park 2-T simulations take much longer to reach equilibrium.

    \subsubsection{High pressure case $\left(\SI{10000}{Pa}, \SI{50}{kW}\right)$}
\begin{figure}[!htb]
    \hspace*{-0.75cm}
    \subfloat[][]{\includegraphics[scale=0.16]{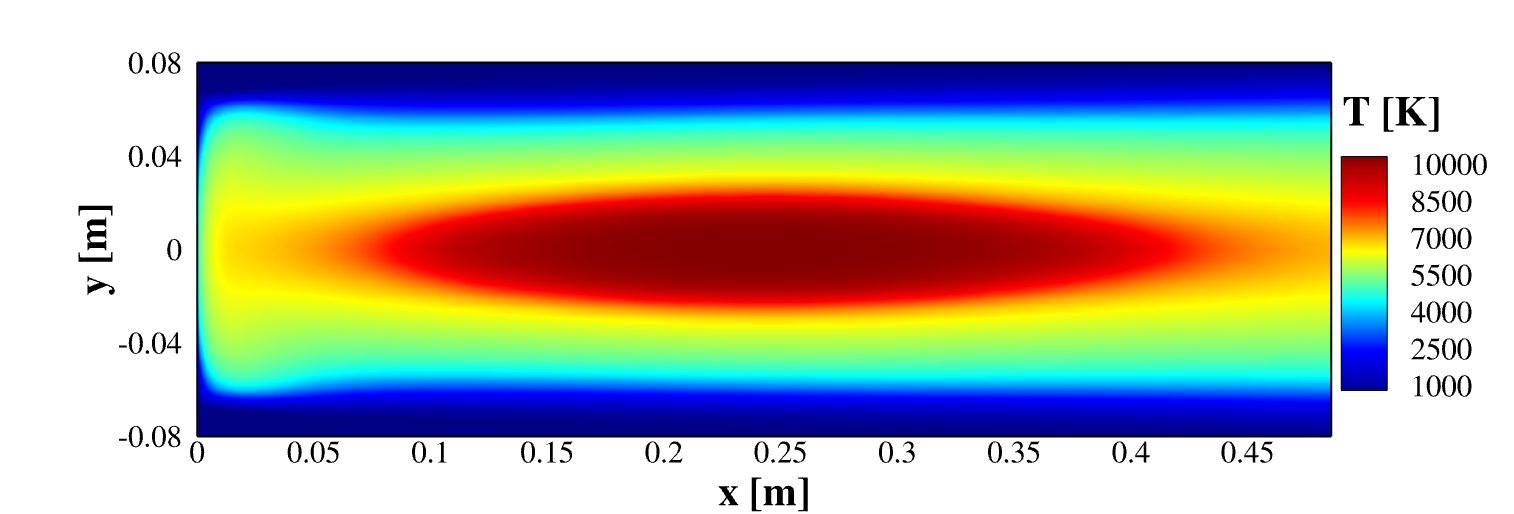}}
    \subfloat[][]{\includegraphics[scale=0.16]{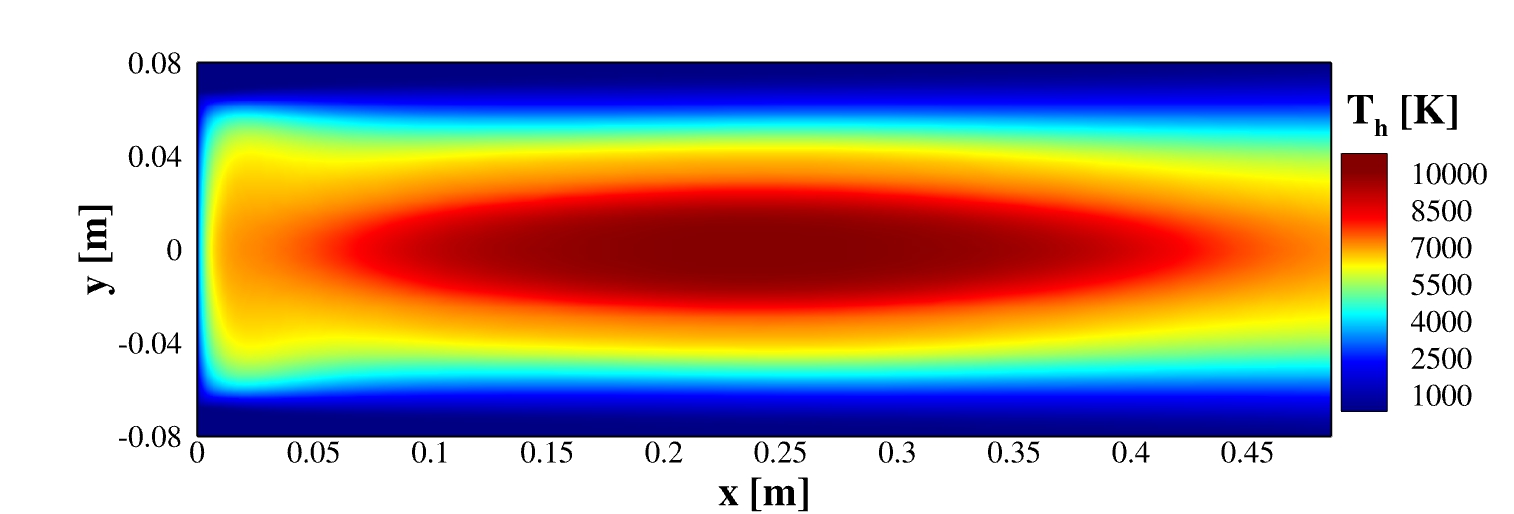}}\\
    \hspace*{-0.75cm}
    \subfloat[][]{\includegraphics[scale=0.16]{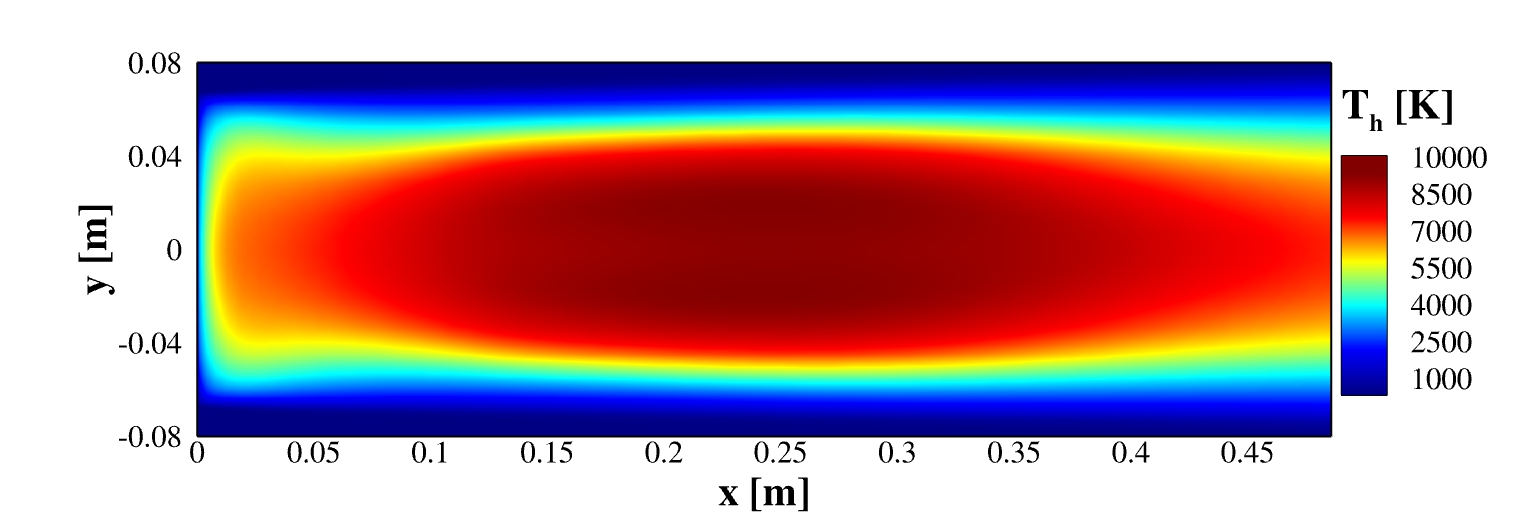}}
    \subfloat[][]{\includegraphics[scale=0.16]{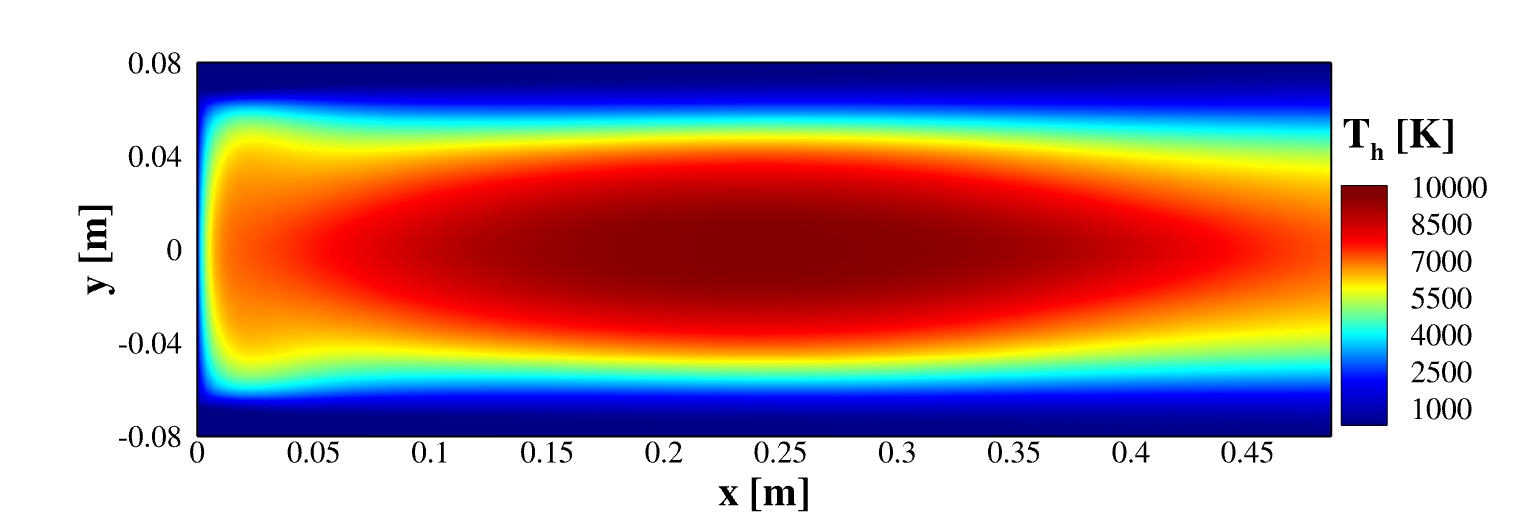}}
    \caption{Plasma heavy-species temperature contours: (a) LTE, (b) Park 2-T, (c) consistent 2-T, and (d) vibronic StS. Operating conditions: \SI{10000}{Pa}, \SI{50}{kW} and \SI{6}{g/s}.}
    \label{fig:T_contour_model_compare_10KPa_50KW}
    \end{figure}

\begin{figure}[!htb]
    \hspace*{-0.75cm}
    \subfloat[][]{\includegraphics[scale=0.55]{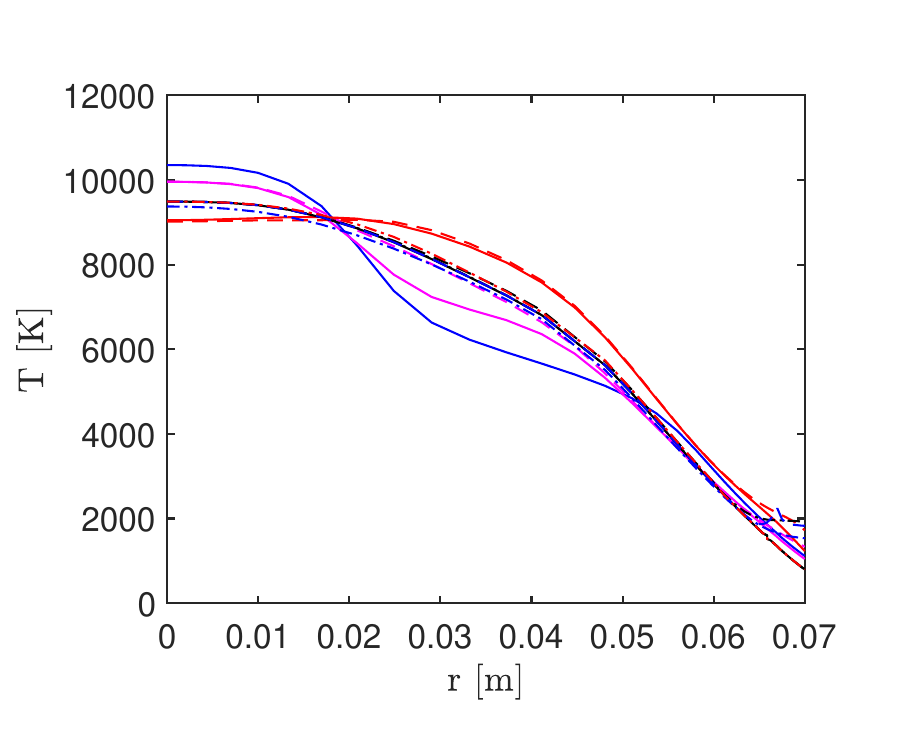}}
    \subfloat[][]{\includegraphics[scale=0.55]{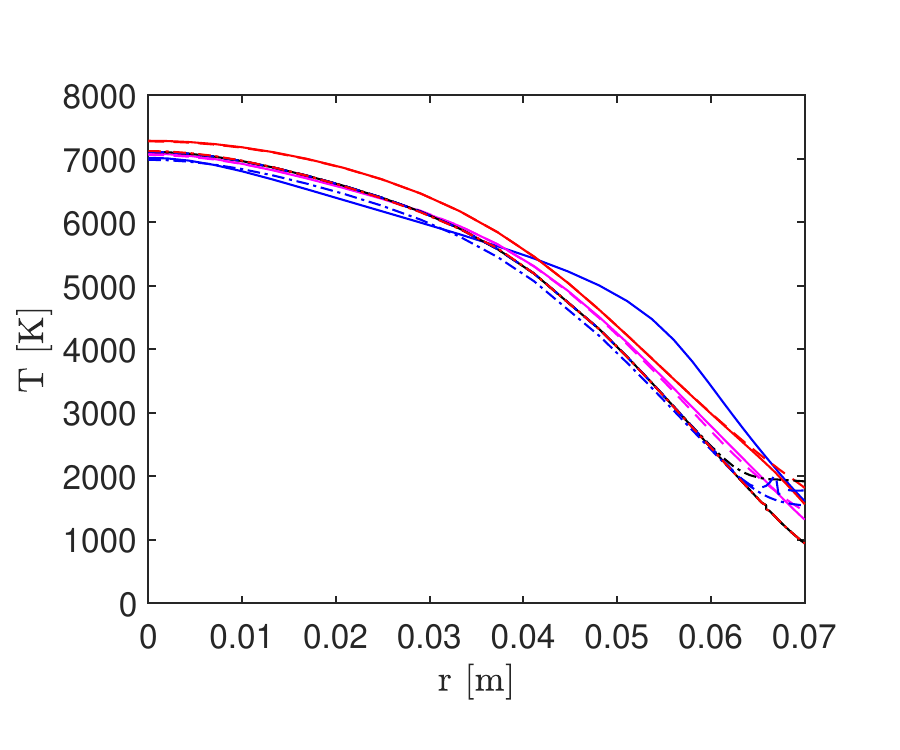}}
    \caption{Radial temperature profiles at: (a) x = \SI{0.3}{m} (mid-torch location) and (b) x = \SI{0.485}{m} (torch outlet). Solid blue line: T (LTE), solid magenta line: T\textsubscript{h} (Park 2-T), dashed magenta line: T\textsubscript{ev} (Park 2-T), solid red line: T\textsubscript{h} (consistent 2-T), dashed red line: T\textsubscript{ev} (consistent 2-T), solid black line: T\textsubscript{h} (vibronic StS), dashed black line: T\textsubscript{e} (vibronic StS), dashed-dot black line: T\textsubscript{N} (vibronic StS), dashed-dot blue line: T\textsubscript{N\textsubscript{2}} (vibronic StS), dashed blue line: T\textsubscript{N\textsuperscript{+}} (vibronic StS), and dashed-dot red line: $\mathrm{T}_{\mathrm{N}_2^+}$. Operating conditions: \SI{10000}{Pa}, \SI{50}{kW} and \SI{6}{g/s}. }
    \label{fig:T_profiles_model_compare_10KPa_50KW}
    \end{figure}

    As the pressure inside the ICP torch increases, higher collision frequency allows thermal equilibrium between the electrons and heavy-species. As a result, LTE conditions start to prevail even in the coil region of the torch and hence simulations performed using LTE assumption give reasonably accurate plasma flow field. Moreover, simulations performed using NLTE models should give LTE flow fields at high operating pressures. \cref{fig:T_contour_model_compare_10KPa_50KW} shows the plasma temperature contours inside the torch which are qualitatively similar for all the physico-chemical models, with LTE contour slightly different from the NLTE contours. \cref{fig:T_profiles_model_compare_10KPa_50KW} shows that all the NLTE models (\textit{i.e.} Park 2-T, consistent 2-T and vibronic StS) show thermal equilibrium between the translational and the electro-vibrational modes in both the coil region as well as away from the coils (\textit{i.e.} at the outlet). However, the temperature profiles given by different models although very close, do not overlap exactly with one another and with the LTE temperature profile, especially in the coil region. This indicates that complete LTE conditions do not prevail in the coil region even at this pressure for nitrogen plasma. However, at the torch outlet, the temperature profiles obtained from various NLTE models are close to that of LTE. This indicates that the use of the LTE model above this pressure should be able to give reasonably accurate results in the chamber region of the ICP facility, although pressures above \SI{30}{kPa} is better suited for LTE model as concluded in Part I of this work. For vibronic StS simulation, the vibronic temperature profiles of all the components ($\mathrm{N},\mathrm{N}^+,\mathrm{N}_2,\mathrm{N}_2^+$) collapse with free-electron temperature indicating negligible non-Boltzmann effect. This observation is physically consistent with the fact that at higher pressures, there will be smaller non-equilibrium and non-Boltzmann effect. 
    
    \subsubsection{High power case $\left(\SI{1000}{Pa}, \SI{250}{kW}\right)$ }
\begin{figure}[!htb]
    \hspace*{-0.75cm}
    \subfloat[][]{\includegraphics[scale=0.16]{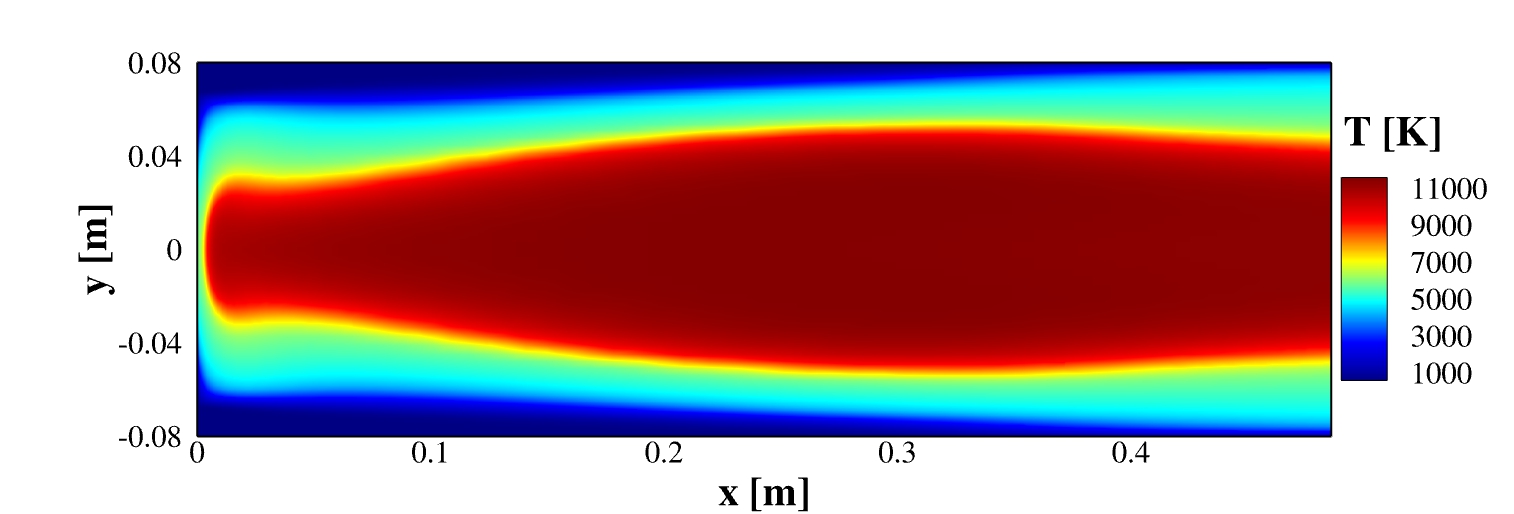}}
    \subfloat[][]{\includegraphics[scale=0.16]{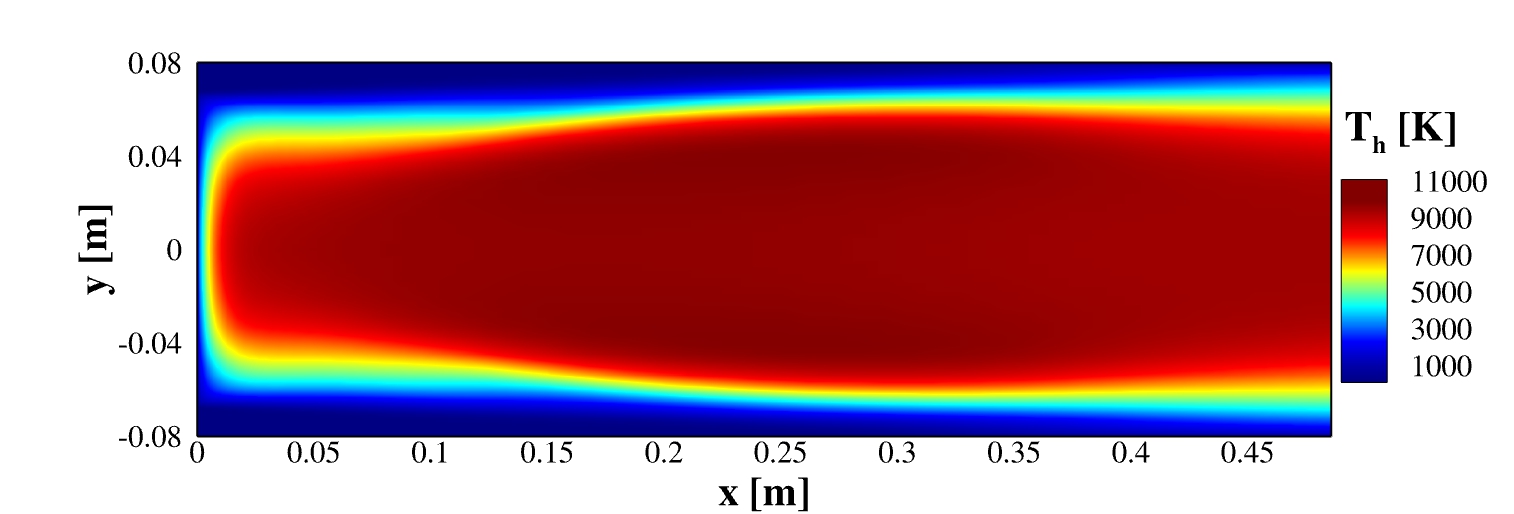}}\\
    \hspace*{-0.75cm}
    \subfloat[][]{\includegraphics[scale=0.16]{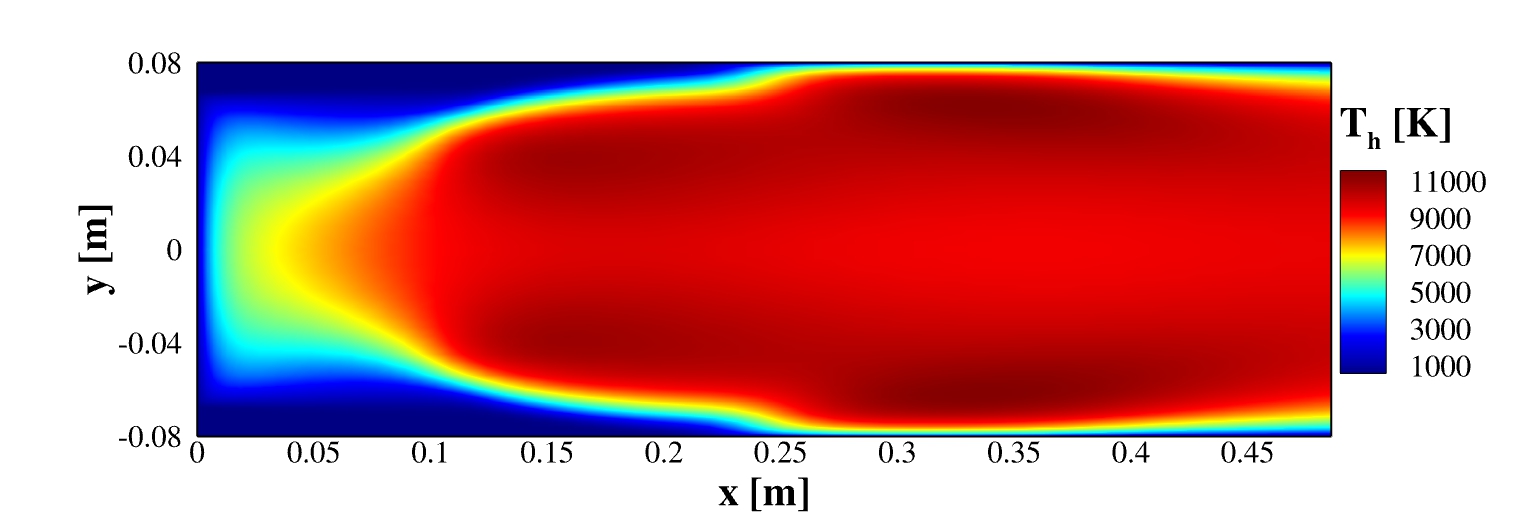}}
    \subfloat[][]{\includegraphics[scale=0.16]{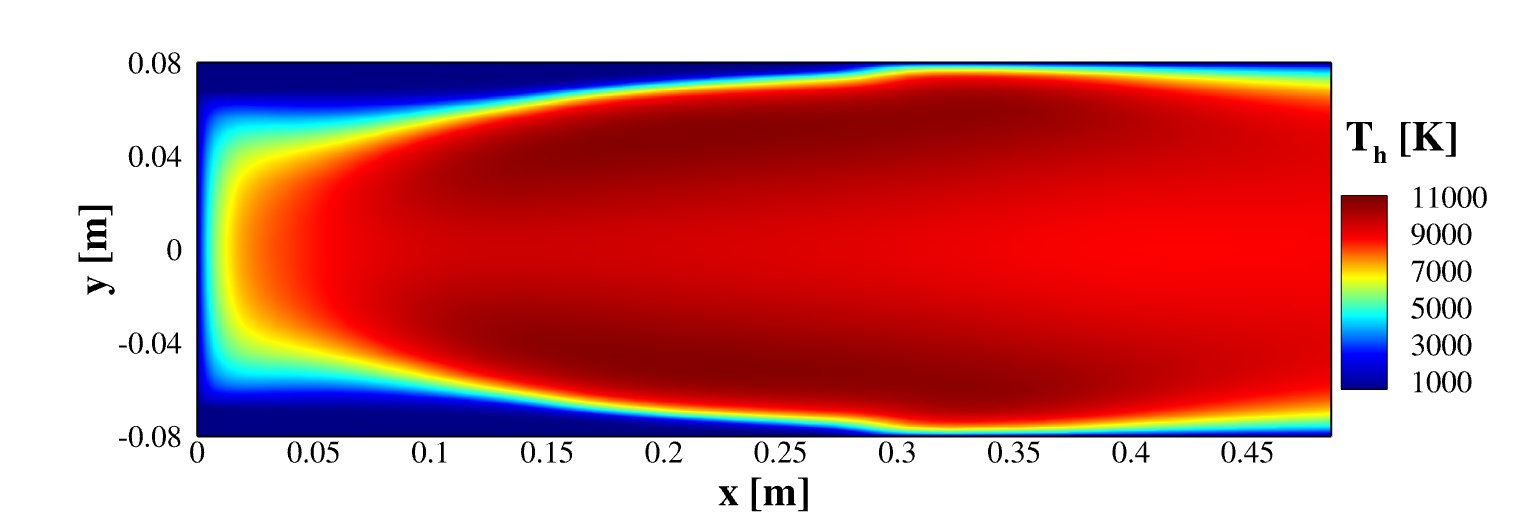}}
    \caption{Plasma heavy-species temperature contours: (a) LTE, (b) Park 2-T, (c) consistent 2-T, and (d) vibronic StS. Operating conditions: \SI{1000}{Pa}, \SI{250}{kW} and \SI{6}{g/s}.}
    \label{fig:T_contour_model_compare_1KPa_250KW}
    \end{figure}

\begin{figure}[!htb]
    \hspace*{-0.75cm}
    \subfloat[][]{\includegraphics[scale=0.55]{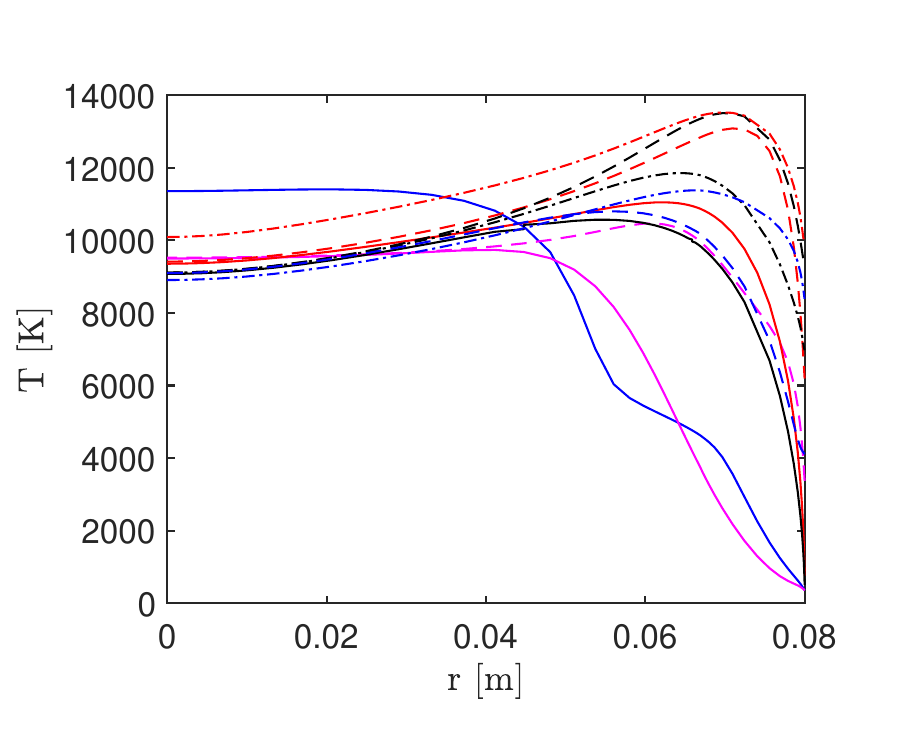}}
    \subfloat[][]{\includegraphics[scale=0.55]{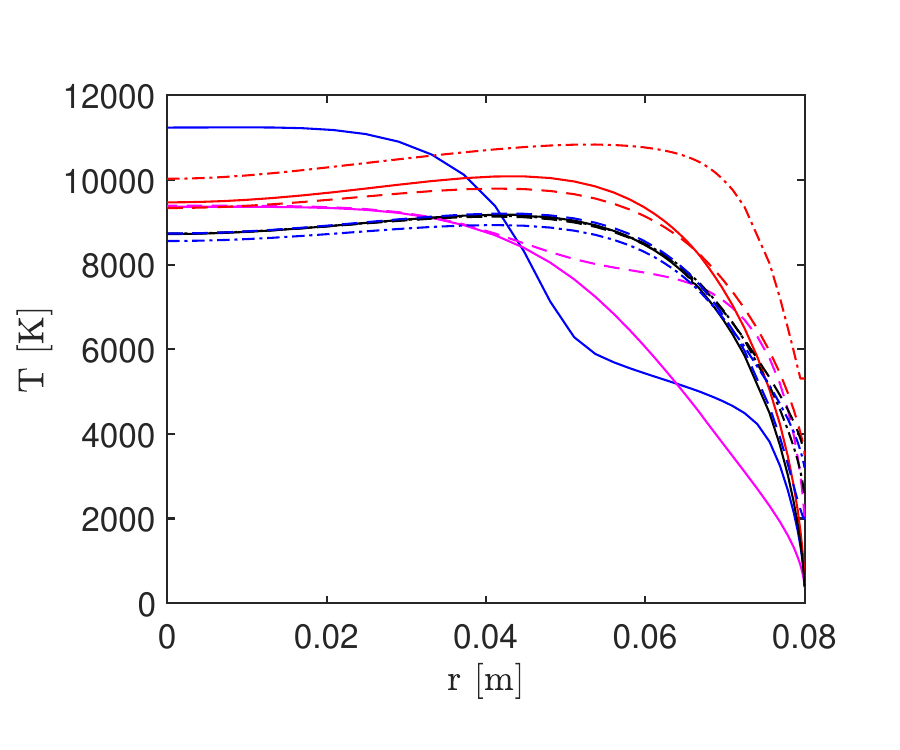}}
    \caption{Radial temperature profiles at: (a) x = \SI{0.3}{m} (mid-torch location) and (b) x = \SI{0.485}{m} (torch outlet). Solid blue line: T (LTE), solid magenta line: T\textsubscript{h} (Park 2-T), dashed magenta line: T\textsubscript{ev} (Park 2-T), solid red line: T\textsubscript{h} (consistent 2-T), dashed red line: T\textsubscript{ev} (consistent 2-T), solid black line: T\textsubscript{h} (vibronic StS), dashed black line: T\textsubscript{e} (vibronic StS), dashed-dot black line: T\textsubscript{N} (vibronic StS), dashed-dot blue line: T\textsubscript{N\textsubscript{2}} (vibronic StS), dashed blue line: T\textsubscript{N\textsuperscript{+}} (vibronic StS), and dashed-dot red line: $\mathrm{T}_{\mathrm{N}_2^+}$ (vibronic StS). Operating conditions: \SI{1000}{Pa}, \SI{250}{kW} and \SI{6}{g/s}.}
    \label{fig:T_profiles_model_compare_1KPa_250KW}
    \end{figure}

    \cref{fig:T_contour_model_compare_1KPa_250KW} shows the plasma temperature contours for very high power cases which leads to high temperatures very close to the top (cold) wall. Since the pressure is quite low, the NLTE flow field is very different from the LTE flow field as seen in the contours. \cref{fig:T_profiles_model_compare_1KPa_250KW} (a) shows significant non-equilibrium between the translational and the electro-vibrational modes especially at around r = \SI{0.07}{m} which is much closer to the top wall as compared to the base case where the peak non-equilibrium was seen at around r = \SI{0.05}{m}. This shifting of the peak non-equilibrium region towards the top wall occurs due to the concentration of Joule heating closer to the wall in case of high power conditions since the inductive heating from the coils is the source of the NLTE effect in the torch. It can be seen that the temperature profiles obtained from Park 2-T are close to vibronic StS near the axis but start deviating as we move closer to the cold wall where a large non-equilibrium effect is seen. The consistent 2-T profiles, however, are close to the vibronic StS profiles, affirming that the consistent 2-T model can reproduce StS results even in this operating condition. The vibronic temperatures of various components ($\mathrm{N},\mathrm{N}^+,\mathrm{N}_2,\mathrm{N}_2^+$) show a large deviation from electron temperature confirming significant non-Boltzmann effect.

\section{Conclusions}\label{sec:conclusions}
    This paper presents a comparative study of various physico-chemical models used for inductively coupled plasma simulations. NLTE ICP simulations using the vibronic StS model described in Part I of this work have been compared against the most widely used Park 2-T model simulations, showing large discrepancies in the plasma flow field. Further, a consistent 2-T model has been developed by reducing the vibronic StS model under QSS assumption, which can reproduce the qualitative characteristics (plasma core location and temperatures) of the StS plasma flow field with much lower computational cost. It was also found that the vibrational relaxation time is a key parameter that controls the plasma core morphology (location and temperature).  However, the consistent 2-T model still gives slight discrepancies in terms of electron concentrations and electron temperature as a result of the 2-T model's inadequacy in accurately modeling phenomena like non-preferential ionization, electron-impact electronic excitation, dissociation, and others. Further, simulations are presented for a range of operating conditions using various models (LTE, 2-T, and vibronic StS) to assess the applicability of various models given the ICP facility operating conditions. Future work will focus on improving the state-to-state as well as the consistent 2-T model by comparison against experiments.

\section*{Acknowledgments}
   This work is funded by the Vannevar Bush Faculty Fellowship OUSD(RE) Grant No: N00014-21-1-295 with M. Panesi as the Principal Investigator. 

\section*{Competing interest}
The authors declare no competing interests.
    
\section*{Appendix}
\appendix
    \begin{appendices}
    \section{Macroscopic reaction rates: Park 2-T versus consistent 2-T model (developed in the present work)}\label{appendix:park_vs_elio_rates}

\begin{figure}[H]
    \hspace*{-1cm}
    \subfloat[][]{\includegraphics[scale=0.4]{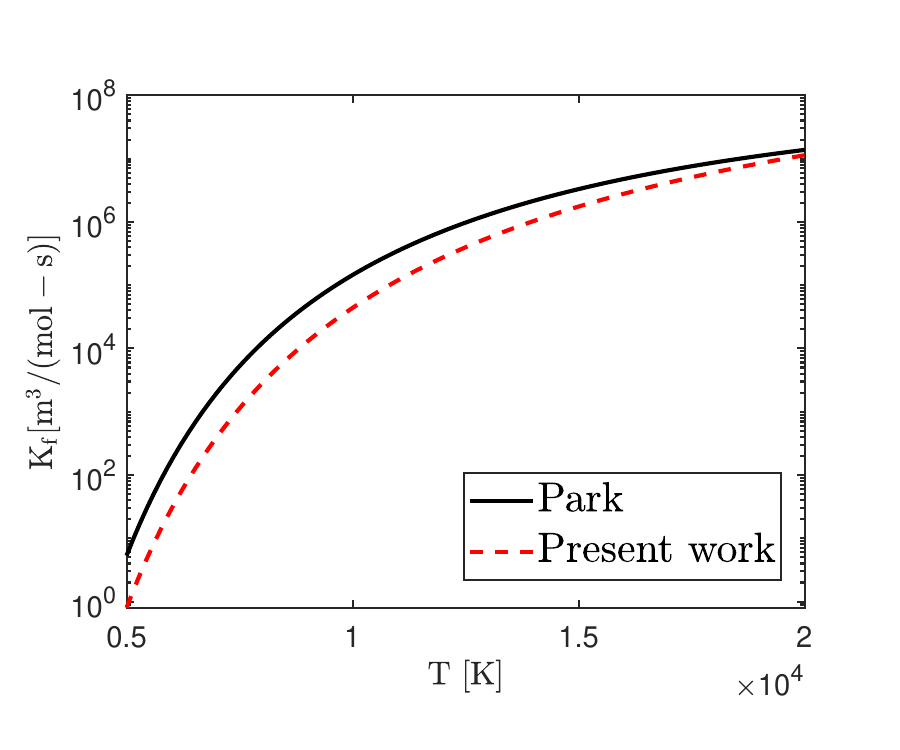}}
    \subfloat[][]{\includegraphics[scale=0.4]{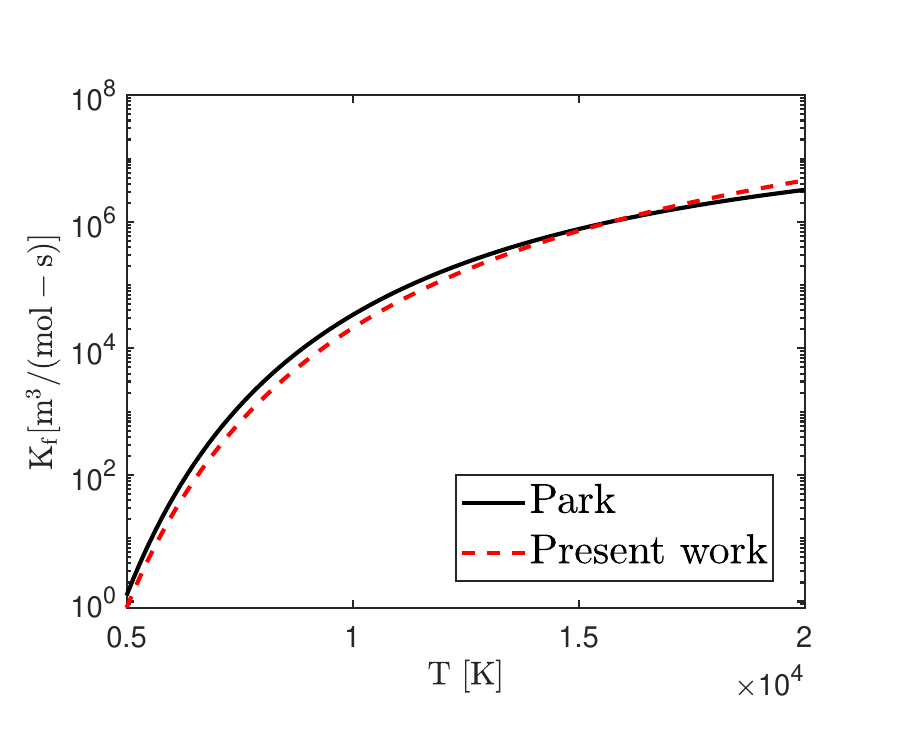}}
    \subfloat[][]{\includegraphics[scale=0.4]{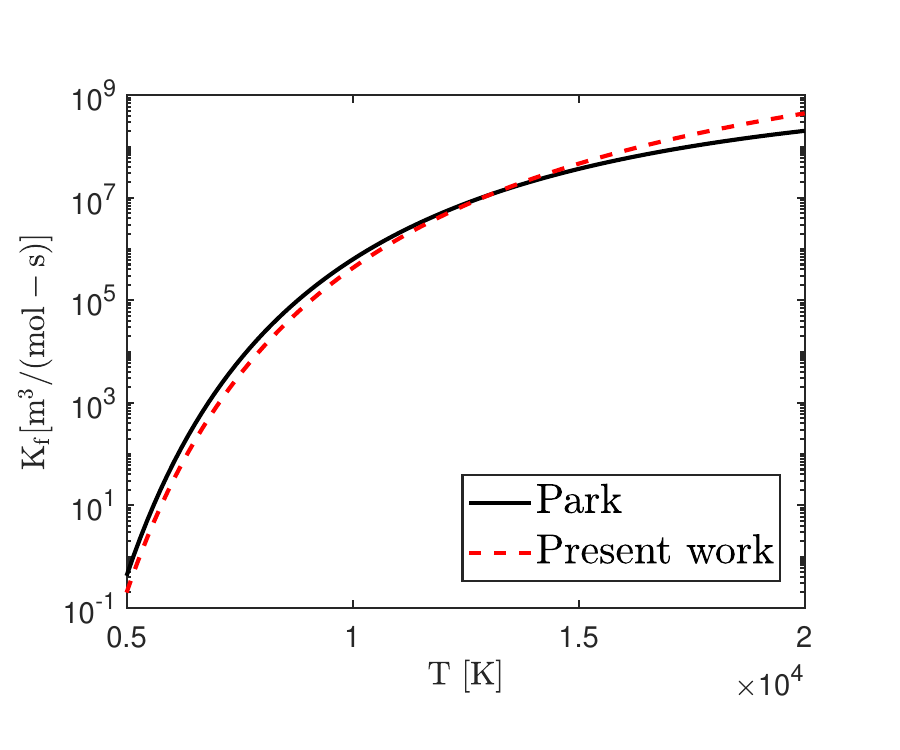}}\\ 
    \subfloat[][]{\includegraphics[scale=0.4]{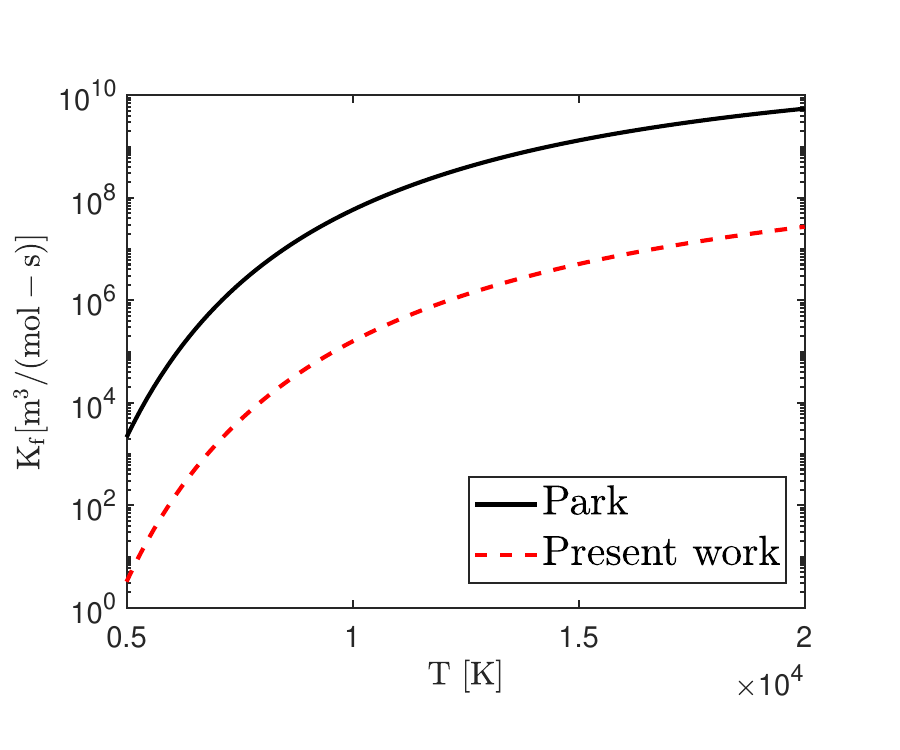}}
    \subfloat[][]{\includegraphics[scale=0.4]{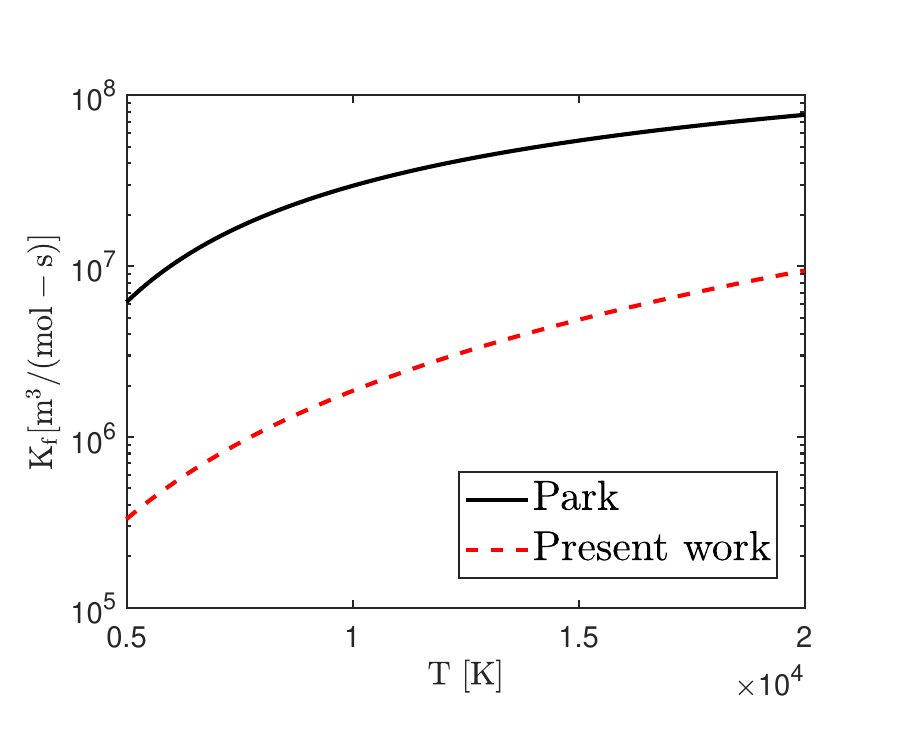}}
    
    \caption{Macroscopic reaction rates: (a) $\mathrm{N}_2 + \mathrm{N} = 3\mathrm{N}$, (b) $\mathrm{N}_2 + \mathrm{N}_2 = 2\mathrm{N} + \mathrm{N}_2$, (c) $\mathrm{N} + \mathrm{e} = \mathrm{N}^+ + 2\mathrm{e}$, (d) $\mathrm{N}_2 + \mathrm{e} = 2\mathrm{N} + \mathrm{e}$, and (e) $\mathrm{N}_2 + \mathrm{N}^+ = \mathrm{N}_2^+ + \mathrm{N}$.}
    \label{fig:park_vs_elio_rates}
    \end{figure}

    This section presents a comparison of the macroscopic rates obtained by reducing the vibronic StS kinetics against Park's macroscopic rates as shown in \cref{fig:park_vs_elio_rates}. 

    \end{appendices}
    
\section*{References}
\bibliographystyle{aiaa}
\bibliography{bibliography}

\end{document}